# Anisotropic Thermal Conductivity of 3D Printed Graphene Enhanced Thermoplastic Polyurethanes Structure toward Photothermal Conversion

Zihao Kang[a], Min Xi*[b], Nian Li[b], Shudong Zhang[b], Zhenyang Wang*[b]

[a] School of Energy, Materials and Chemical Engineering, Hefei University, Hefei, 230601, P. R. China

[b] Institute of Solid State Physics and Key Laboratory of Photovoltaic and Energy Conservation Materials, Hefei Institutes of Physical Science, Chinese Academy of Sciences, Hefei, Anhui 230031, P. R. China

Corresponding authors:

Min Xi − orcid.org/0000-0003-4414-3110; Email: minxi@issp.ac.cn

Zhenyang Wang − orcid.org/0000-0002-0194-3786; Email: zywang@iim.ac.cn

**Table of Contents**

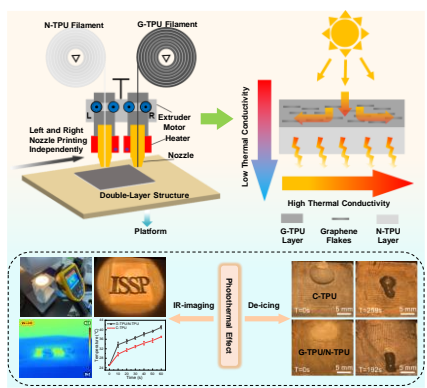

Graphene enhanced thermoplastic polyurethanes (G-TPU) and neat thermoplastic polyurethanes (N-TPU) were developed into double-layer structure via fused deposition modelling (FDM) 3D printing, which exhibited anisotropic thermal conductivity ($TC_{IP}/TC_{TP}$) of ~8 simultaneously satisfied high in-plane (IP) and low through-plane (TP) thermal conductivity, contributed to uniform heat distribution, heat retention, and photothermal conversion, was promising for photothermal de-icing and infrared labels applications.

**Abstract**

Solar photothermal conversion is one of the most straightforward methods to utilize solar energy. In this manuscript, a novel double-layer structure constructed of graphene enhanced thermoplastic polyurethanes (G-TPU) and neat thermoplastic polyurethanes (N-TPU) was developed via fused deposition modelling (FDM) 3D printing process. The developed G-TPU/N-TPU double-layer structure exhibited anisotropic thermal conductivity that simultaneously satisfied high in-plane (IP) thermal conductivity and low through-plane (TP) thermal conductivity. The top G-TPU layer essentially offered a high IP thermal conductivity of 4.54 W/(m·K) that lead to overall structure's anisotropic thermal conductivity ratio (TC$_{IP}$/TC$_{TP}$) of ~8. And the low thermal conductivity in the TP direction led to the heat retention effects for thermal storage. Nonetheless, the exceptional photothermal conversion effect of graphene flakes guaranteed the superior photothermal performance that was promising in the photothermal de-icing and infrared labels applications. Finally, the graphene flake's enhancement in the mechanical properties of the G-TPU/N-TPU double layer structure was also evaluated that contributed to excellent impact resistance with a puncture energy reaching 12.86 J, and extraordinary wear resistance with a small friction coefficient of 0.1 over 1000 cycles, which ensured the structure suitable for applications at harsh environment.

**Introduction**

Solar photothermal conversion has been widely used in various applications, such as: solar steam generation (SSG),[1] anti-icing/de-icing performance,[2] cell resuscitation and rewarming,[3] etc. In particular, recent studies have proposed the integration of physically unclonable functions (PUFs) with unique optical properties, that the optical response generated upon illumination is unique and unpredictable, varying with the viewing angle.[4,5] Mazzotta et al building on PUF system proposed an anti-counterfeiting approach based on the direct printing of a plasmonic ITO NP-based custom ink to obtain arbitrary patterns with high resolution, which are invisible to the naked eye but evident when exposed to near-infrared (NIR) radiation.[6] Cheng et al developed a highly flexible and stable plasmonic nanopaper (PNP) composed of AgNCs and CNFs for unclonable anti-counterfeiting labels, leveraging Surface-Enhanced Raman Scattering (SERS) mapping-based PUF encryption.[7] The developed PUF security labels outperformed other traditional anti-counterfeiting technologies, such as digital watermarking,[8] holograms,[9,10] barcodes,[11,12] and luminescence printing, in terms of information leakage and duplication susceptibility.

The key to improve the photothermal performance is to generate heat localization, that is to simultaneously fulfill high photothermal conversion rate and avoid heat dissipation. Graphene's unique electronic band structure enables effective absorption across a broad spectrum of light, making it has high efficiency of photothermal conversion.[13,14] At the same time, graphene flakes exhibit excellent in-plane (IP) thermal conductivity, ranging from 1000 to 3000 W/(m·K), while their through-plane (TP) thermal conductivity is only approximately 5 W/(m·K),[15] and this high anisotropic thermal conductivity ratio endows graphene with exceptional directional heat transfer capabilities, which helps with the uniform thermal distribution within the surface.[16] These extraordinary thermal properties make graphene a perfect candidate for photothermal conversion.

To fully utilize the anisotropic thermal properties of graphene flakes, different assembly methods have been tried to assembly graphene flakes in the nanocomposites, such as freeze drying,[17] template methods,[18] electrical or magnetic field assistance,[19,20] three-dimensional (3D) printing,[21,22] etc. Specifically, the extrusion-based 3D printing method, which is a straightforward and convenient approach involving the extrusion of thermoplastic filaments, employs shear force to induce alignment of graphene flakes in the preferred direction within the laminar flow system of melted polymer during the fused deposition modelling (FDM) 3D printing process.[23,24]

In this study, a double-layer structure comprising graphene enhanced thermoplastic polyurethanes (G-TPU) and neat thermoplastic polyurethanes (N-TPU) was realized through a convenient and economical approach as 3D printing. The structure was designed and fabricated using a layer-by-layer assembly process with a double-nozzle FDM 3D printer. The printed G-TPU/N-TPU double-layer structure had exceptional anisotropic thermal conductivity of ~8 that simultaneously satisfied high IP thermal conductivity and low TP thermal conductivity due to the alignment of graphene flakes during the FDM printing process of G-TPU layer and the thermal insulation nature of N-TPU layer. Further studies on the directional thermal conductivity and heat retention effects of G-TPU/N-TPU double-layer structure were also conducted and supported by numerical simulations. The IP thermal conductivity of the G-TPU composite in the printing direction was recorded at 4.54 W/(m·K) that was ~6 times higher than that of TP direction, which contributed to anisotropic thermal conductivity of ~8 in the G-TPU/N-TPU double-layer structure that was further improved by the thermal insulated N-TPU of bottom layer. And the G-TPU/N-TPU double-layer structure showed an equilibrium temperature that was ~10 °C higher than that of the G-TPU single-layer structure over the same cooling period as the excellent heat retention effect. Moreover, due to the exceptional photothermal conversion effect of graphene flakes, the G-TPU/N-TPU double-layer structure showed superior photothermal performance. Under simulated sunlight irradiation of 0.15 W/cm², the heating rate of the double-layer structure was ~4 °C higher than that of the colored black TPU (C-TPU) material, which was essentially promising for de-icing performance and PUF labels. Furthermore, the mechanical properties of this G-TPU/N-TPU double layer structure was investigated. It demonstrated excellent impact resistance with a puncture energy reaching 12.86 J, and extraordinary wear resistance with a small friction coefficient of 0.1 over 1000 cycles. Supported by molecular dynamic simulations, the introduced graphene flakes induced strong interaction between polymer chains and graphene flakes, that led to enhanced elastic modulus and shear modulus that contributed to the overall improvement of G-TPU/N-TPU double layer structure's mechanical properties. It was proved that the G-TPU/N-TPU double layer structure with anisotropic thermal conductivity for directional heat transfer and insulation and extraordinary photothermal properties was potential for photothermal de-icing and PUF security labels applications.

## Results and Discussion
### 1. Preparation of Filaments and Samples

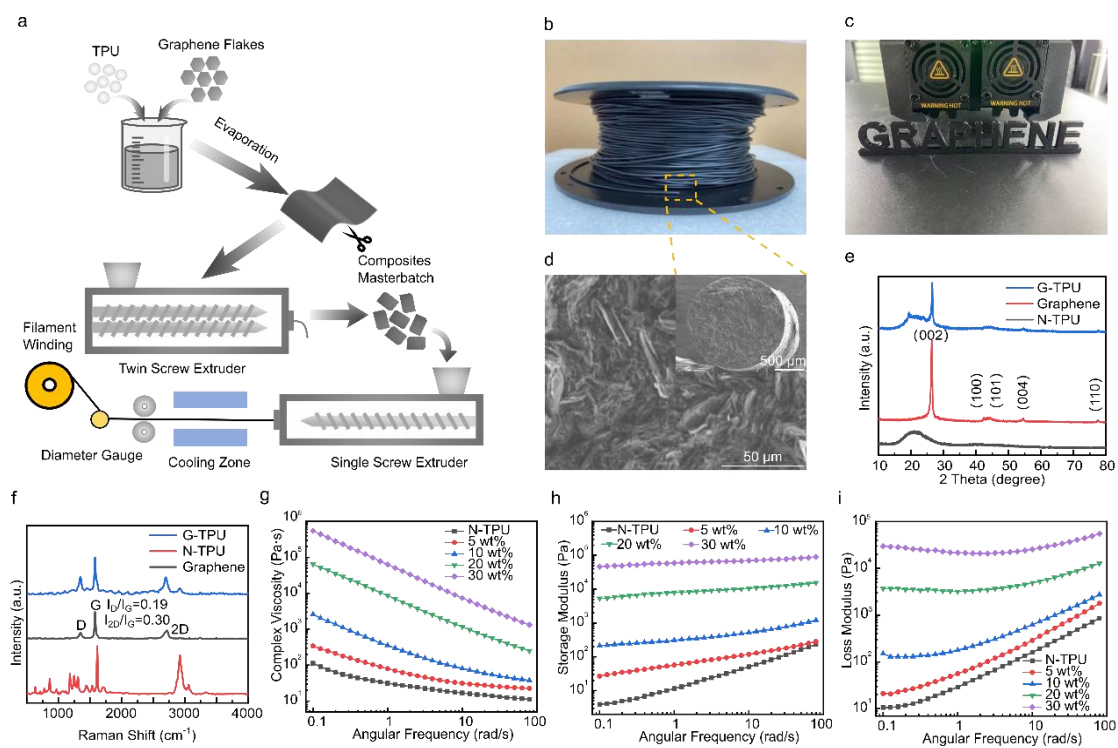

**Figure 1** Synthesis of G-TPU 3D printing filament. **(a)** Schematic illustration of the fabrication of G-TPU filaments; **(b)** The prepared G-TPU filament with diameter of 1.75 mm for FDM 3D printing composed of 30 wt% graphene flakes; **(c)** 3D printed model using prepared 30 wt% G-TPU filament; **(d)** SEM image of cross-sectional morphology of 30 wt% G-TPU filament; **(e)** XRD spectrum of N-TPU, graphene flakes and 30 wt% G-TPU composite filament; **(f)** Raman spectrum of N-TPU, graphene flakes and G-TPU composite filament; **(g) − (i)** Complex viscosity, storage modulus and loss modulus of different G-TPU composite filament as function of scanning frequency.

Figure 1a illustrated the complete fabrication process of 3D-printed G-TPU composite filaments as reported.[25] A two-steps mixing method was applied, encompassing both solution mixing and melt blending to ensure an optimal dispersion of graphene within the TPU matrix, minimizing voids and aggregation.[26,27] Basically, TPU pellets and graphene flakes (**Figure S1**) were thoroughly mixed in DMF solvent, and subsequently, the cured graphene-TPU films were cut into small pieces and fed into a twin-screw extruder for melt blending. The processed masterbatch (**Figure S2**) was then extruded through a single-screw extruder, coupled with a small winding machine, to produce and collect filaments with a diameter of 1.75 mm suitable for FDM printing. **Figure S3** displayed the TGA curves of G-TPU composite filaments and N-TPU filament. The filler content in the composite was calculated by subtracting the residual weight of the N-TPU from the overall weight of the composite.[28] The results were consistent with the weight of the initially added graphene as nominal weight percentage, indicating that there was essentially no loss of filler during the preparation of the composite filament. **Figure 1b** showed the representative filament utilized for FDM, which had a diameter of 1.75 ± 0.1 mm. It was evident that after adding 30 wt% graphene, the filament's surface was smooth and well-formed, and could be easily coiled, collected, and stored for 3D printing purposes. **Figure 1c** displayed an object printed of the prepared 30 wt%

G-TPU filament mentioned above. The sample was in high quality with smooth surface, and the filament extruded continuously from the nozzle throughout the printing process. This demonstrated that the filament possesses excellent printability and the capability to print complex and exquisite structures. **Figure 1d** displayed the micrographs of cross sections of extruded composite filament from the nozzle. The images demonstrated that the graphene flakes were evenly distributed in the TPU matrix with a vortex morphology. And in consistent to the peer reference, the graphene sheets were oriented in the axial direction of the filaments as the shear force during extrusion, which resulted to the anisotropic thermal properties of the printed structure.[25]

The characteristic peaks of (002), (101), and (004) peaks of XRD diffraction patterns in **Figure 1e** confirmed the existence of graphene monolayer and graphene sheets of a few layers, which were devoid of impurities.[2] Purity is critical for enhancing thermal conductivity and photothermal properties, as impurities can act as phonon scattering centers, thereby reducing the thermal conductance of the final composites.[29] Importantly, the spectrum of the G-TPU composites displayed the same diffraction peaks as those of the graphene flakes that suggested that the graphene remained intact during the fabrication process. The Raman spectrum characterization results were presented in **Figure 1f**. The characteristic D peak, G peak, and 2D peak of graphene flakes provided insights into the structural defects, degree of graphitization, and layer number of the graphene, respectively.[30] A low $I_D/I_G$ ratio of 0.19 indicated the integrity of graphene sheets, while an $I_{2D}/I_G$ value of 0.30 suggests the few-layer characteristic of the graphene flakes. The Raman curve of the G-TPU composite appeared as a perfect combination of the TPU and graphene characteristic peaks, indicating the uniform distribution of graphene flakes within the TPU polymer matrix.

**Figure 1g – 1i** showed the changes in complex viscosity, storage modulus (G'), and loss modulus (G") of samples of different graphene weight percentage under scanning frequency at 230 °C, respectively. The results indicated that with increasing frequency, both G' and G" steadily increase, while the complex viscosity decreased due to the shear thinning of the polymer melt.[31] The G-TPU composites exhibited increased modulus and viscosity compared to N-TPU. This was attributed to the reduction in free volume caused by the filler, which lead to the adsorption of macromolecules onto the filler, thereby limiting the movement of the polymer chains. The comparatively low viscosity of the molten material enabled even distribution of graphene.[32] **Figure 1g** exhibited the shear thinning property of the polymer melt, which ensured seamless extrusion. Notably, due to the confinement of graphene flakes by neighboring sheets, the storage modulus (G') of G-TPU composites in **Figure 1h** exhibited a frequency-dependent plateau-like pattern. This plateau-like curve suggested excellent printability for extrusion-based printing.

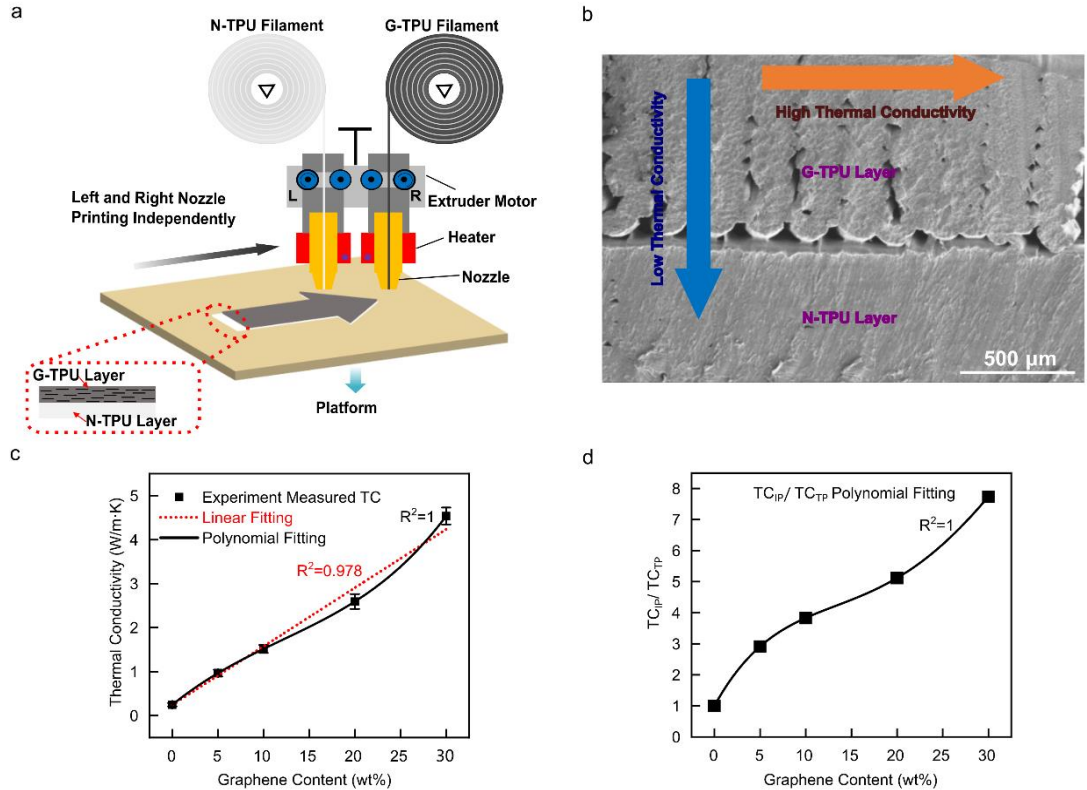

**Figure 2** FDM 3D printed G-TPU/N-TPU double layer structure. **(a)** Schematic illustration of the dual-nozzle FDM 3D printing process; **(b)** SEM image of cross sections of G-TPU/N-TPU double layer structure samples with 30 wt% G-TPU filament and N-TPU filament; **(c)** Thermal conductivity of G-TPU filaments with varying graphene content; **(d)** Anisotropic thermal conductivity ratio of the G-TPU/N-TPU double layer structure with different graphene contents in G-TPU layers.

The as prepared G-TPU filament was subsequently used to assemble into a double-layer structure that was designed to exhibit efficient heat collection and retention properties. The fabrication of this G-TPU/N-TPU double layer structure was achieved through a dual-nozzle 3D printer as illustrated in **Figure 2a**. The FDM printing process initiated with a standard melt extrusion procedure (the printing parameters were listed in **Table S1**), where a 1.75 mm filament was fed into a high-temperature nozzle. The left nozzle employed N-TPU filament, while the right nozzle utilized G-TPU filament. The printing path and stacking manners were configured using slicing software, enabling the left and right nozzles to independently print each layer structure. These layers were then stacked to form a macroscopic 3D entity. **Figure 2b** showed the cross-sectional SEM images of the double-layer structure, where the stacked layer-by-layer structures of G-TPU and N-TPU were clearly observable, confirming the molding characteristics inherent to FDM. As an anisotropy filler, due to the high in-plane thermal conductivity of graphene and its orientation parallel to the printing direction, induced by shear forces in the G-TPU layer, heat flux is rapidly conducted within this layer. While the combination of graphene's through-plane thermal conductivity and the polymer matrix's inherently low thermal conductivity limits heat loss through the double-layer structure. This configuration effectively facilitates directional heat conduction and insulation, optimizing the thermal management capabilities of the printed material.

**Figure 2c** illustrated the in-plane thermal conductivity of G-TPU filament printing samples with varying graphene contents along the printing direction. First, it was evident that the thermal conductivity of all samples increased with the addition of graphene content. This enhancement was attributed to the introduction of thermally conductive fillers. The thermal conductivity of neat TPU was noted to be merely 0.24 W/(m·K). Upon the incorporation of graphene sheets, there was a notable enhancement in thermal conductivity, which peaked at 4.54 W/(m·K) for samples with a graphene content of 30 wt%. Importantly, the increase in thermal conductivity was not linear. Upon analysis, it was found that the variation in thermal conductivity aligned more closely with the results of higher order polynomial fitting. This observation was attributed to the difficulty of thermal conductive fillers to contact with each other and formed effective heat conduction pathways and networks at lower loadings. **Figure 2c** also demonstrated the thermal conductivity of the samples exhibited a sudden increase when the graphene flakes content rises from 20 wt% to 30 wt%. This indicated the formation of a thermal conductive network, which corresponded to the previous studies that optimal thermal conductivity in composites was achieved when the amount of thermal conductive filler was sufficient to form a complete conductive path and network.[33] At this point, the number of interfaces between the larger thermal conductive fillers and the polymer matrix was minimized, which significantly enhanced the thermal conductivity of the composite material.[34]

**Figure 2d** displayed the ratio of overall in-plane (IP) to through-plane (TP) thermal conductivity for G-TPU/N-TPU double-layer structure with varying graphene contents in G-TPU layers (measurement was performed with the samples indicated in **Figure S4a** and **S4b**). Similar to the improved thermal conductivity in **Figure 2c**, the anisotropic thermal conductivity increased with the graphene flake contents. Notably, the ratio of IP to TP thermal conductivity for the sample printed with a filament containing 0 wt% graphene (i.e. N-TPU) was 1, indicating the isotropic thermal properties of TPU. In contrast, the anisotropic thermal conductivity ratio for the sample printed with 30 wt% G-TPU filament was 7.87, demonstrating that the designed double-layer structure model possessed effective directional heat transfer and heat retention properties.

2. Anisotropic Thermal Conductivity Analysis

Next, the thermal conductivity of 3D printed layers was studied. Especially, the stacking manners of extruded G-TPU filament during the printing were considered, where the graphene flakes aligned parallel to the flow direction due to the shear force exerted by the inner wall of the nozzle, resulting in a shearing effect. The shear force ensured the alignment of graphene within the filament during the stacking, while friction with the nozzle might induce the deposition of horizontally arranged graphene sheets atop the filament.[35] It was interesting that the orientation of aligned graphene flakes would lead to anisotropic enhancement in polymer composites mechanical, thermal and optical properties. In our case, the high thermal conductivity in printing direction and low thermal conductivity in stacking direction further explained the anisotropic thermal conductivity of G-TPU/N-TPU double layer.

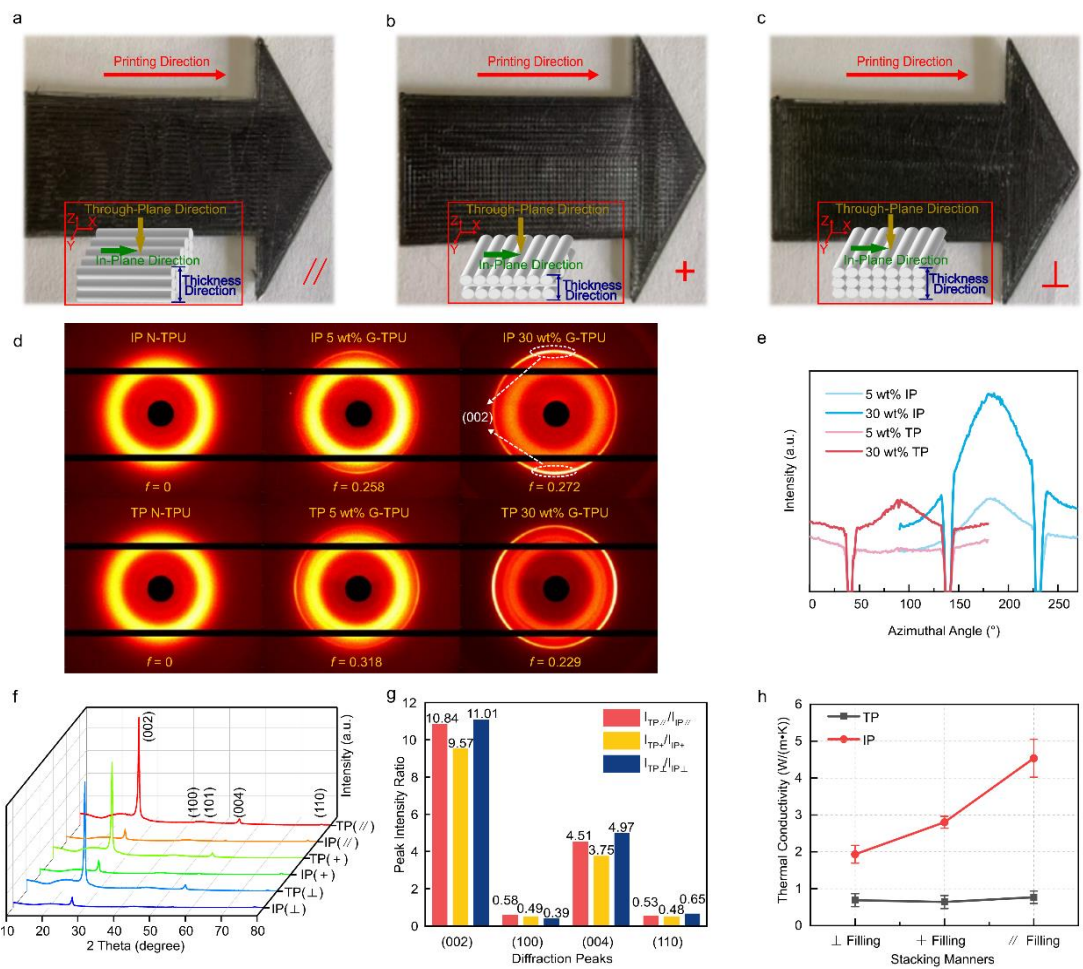

**Figure 3** Characterization of anisotropic G-TPU layer with different stacking manners. G-TPU layer structures printed in different stacking manners: **(a)** [0°/0°]; **(b)** [0°/90°]; **(c)** [90°/90°], where in-plane (IP) and through-plane (TP) directions were indicated in the figures; **(d)** 2D WAXS patterns of N-TPU, 5 wt% G-TPU, and 30 wt% G-TPU samples oriented in the TP and IP directions; **(e)** Azimuthal angle-integrated intensity profile curves of the 2D scattering patterns; **(f)** Comprehensive XRD patterns of the composite samples printed with various stacking manners by 30 wt% G-TPU filament, from IP and TP directions; **(g)** The ratio of characteristic diffraction peak intensities of graphene flakes in XRD patterns of **Figure 3f**; **(h)** Thermal conductivity of printed G-TPU layer with different stacking manners.

To study the analyze the influence of stacking manners on thermal conductivity, shown in **Figure S5**, the anisotropic thermal conductivity of 30 wt% G-TPU layer was revealed as a high thermal conductivity of 4.54 W/(m·K) in the printing direction and a low thermal conductivity of 0.77 W/(m·K) in thickness direction. Moreover, as depicted in **Figure 3a – 3c** and **Figure S6**, three types of double-layer arrow samples were printed using 30 wt% G-TPU filament employing different stacking manners. The figure also illustrated the differences in various stacking manners, as well as in-plane and through-plane directions. XRD patterns were commonly used to estimate the orientation of graphene in the polymer matrix.[36] Quantitative analysis of the graphene flake orientation during the 3D printing process was conducted using 2D WAXS as shown in **Figure 3d**. The G-TPU samples with different graphene flakes loadings in the TP and IP directions displayed an arc shape, which stood in stark contrast to the isotropic rings of the TPU samples. The outer halo corresponded to the (002) plane of graphene, revealing the stacking orientation characteristics of the graphene. The arc shape intensified with the increasing graphene content in the samples. The azimuthal integral intensity of 2D scattering patterns with different graphene contents was shown in **Figure 3e**. The integral intensity significantly increased with the graphene content. The orientation parameter *f* of graphene flakes was determined by calculating the azimuthal angle and intensity of the (002) crystal plane, as indicated in **Figure 3d**. The orientation parameter was highest for the TP sample with 5 wt% graphene content and lowest for the TP sample with 30 wt% graphene content. This was because when the filler content was high, the composite melt exhibited higher viscosity. Consequently, the graphene flakes experience greater viscous resistance during the extrusion orientation process, which reduces the effectiveness of the compression effect during deposition.[37]

**Figure 3f** displayed the XRD patterns of samples printed using 30 wt% G-TPU filament with varying stacking manners in both IP and TP directions. To be more specific, the samples exhibited strong (002) and (004) peaks in the TP direction for each stacking manner, indicating exposure of these crystal faces to X-ray irradiation. This suggested that the majority of the graphene flakes were oriented horizontally within the sample, aligned with the print direction.[25] Comparatively, in the IP direction, all different stacking manner samples exhibited attenuation of the (002) peak and near disappearance of the (004) peak, while the intensity of the (100) and (110) peaks increased. The peaks observed at 26.4° and 54.5° in the XRD patterns could be attributed to the (002) and (400) crystal planes of graphene, respectively. These crystal planes represented the TP orientation of graphene. While the diffraction peak at 42.2° and 77.9° represents graphene (100) and (110) crystal plane, which was the IP direction and the (100) and (110) lattice planes were perpendicular to the (002) and (004) crystal faces,[38] this result further confirmed that the majority of the graphene flakes in the composite are flat-aligned in the XY plane and oriented parallel to the printing direction during the printing process just as expected. Additionally, the (101) peak at 44.9° reflecting the rotation or translation variation of graphene flakes and the consistent plateau-like peaks near 20.8° across all samples reflected the amorphous nature of TPU.[39] To measure the degree of orientation of filler in various samples, the alignment of the graphene flakes within the sample was assessed by analyzing the intensity ratio of graphene diffraction peaks in different directions.[40] As illustrated in **Figure 3g,** the ratio of peak intensities for the (002) crystal face between the through-plane and in-plane directions reached a maximum of 11. Additionally, the maximum characteristic peak intensity ratio for the (004) crystal plane was 4.97. In contrast, the peak intensity ratios for the (100) and (110) crystal faces in both directions were considerably lower, with the minimum value being only 0.39. These observations indicated that: (1) graphene was highly oriented along the printing direction across all samples; (2) different stacking manners did not significantly influence the orientation of the graphene layers. In summary, these findings indicate that graphene flakes in the printed samples exhibited a certain degree of orientation, which aligned with the expected printing direction.

**Figure 3h** shows comparison of thermal conductivity in IP direction and TP direction of 30 wt% G-TPU printed samples with different stacking manners. The thermal conductivity in TP direction exhibits lower values than those in IP direction for all three stacking cases, this finding of anisotropic thermal conductivity in the 3D printing polymer composite corresponded to other research groups.[25,41] This phenomenon could be attributed to the high in-plane orientation of graphene flakes, which aligned parallel to the printing direction, thereby forming an effective heat channel for transporting heat flux.[42]

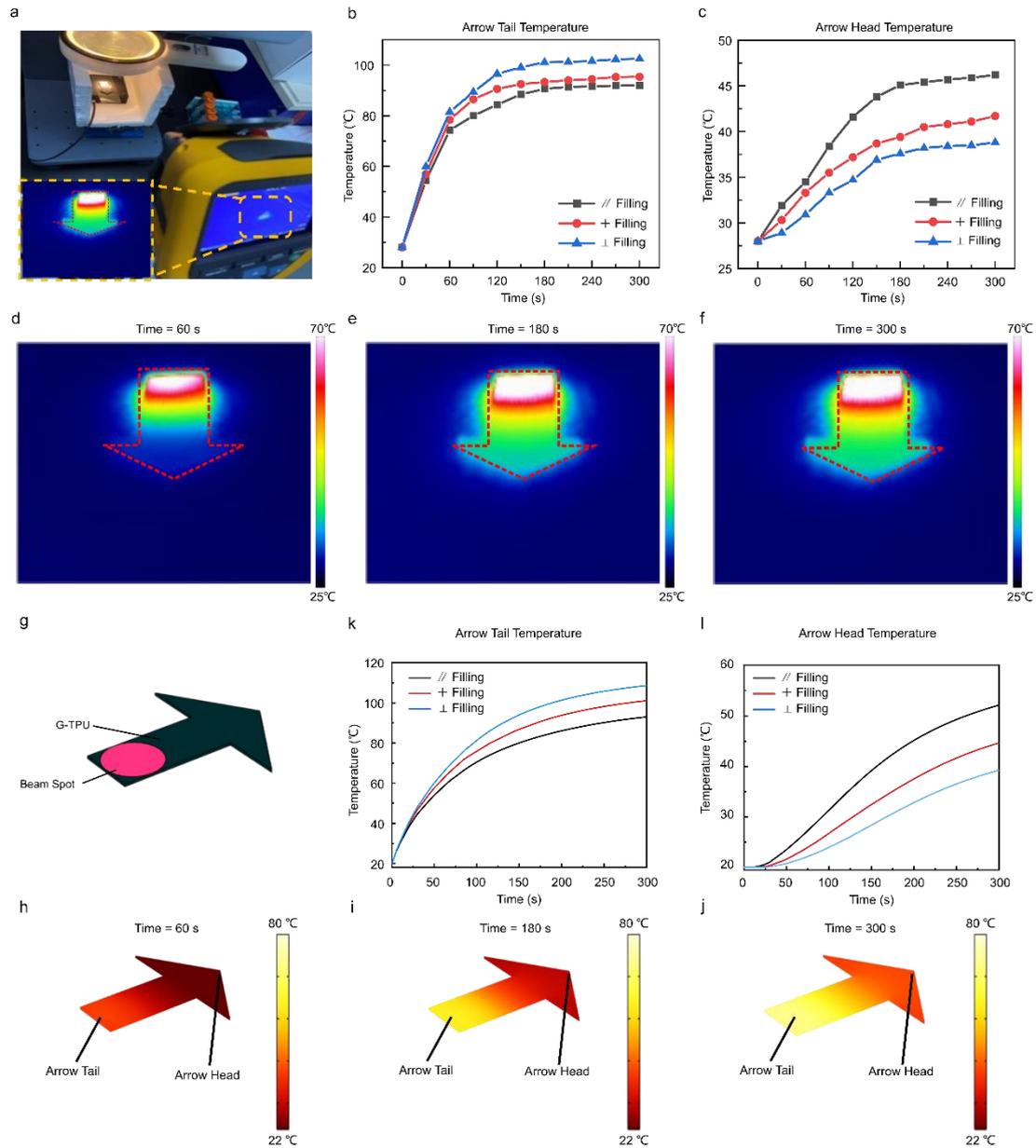

**Figure 4** Anisotropic heat transfer evaluation of G-TPU layer structures with different stacking manners. **(a)** Photograph of the experiment setup for thermal conductivity testing; Temperature changes at the **(b)** tail and **(c)** head of arrow sample under converged solar illumination as function of time; **(d)** − **(f)** Representative infrared images of arrow samples printed in [0°/90°] stacking manner, irradiated by a simulated solar light source at 60 s, 180 s, and 300 s respectively; **(g)** Simulation design; **(h)** – **(j)** Representative calculated temperature distribution of G-TPU layer arrow samples printed in [0°/90°] stacking manners, irradiated by a simulated solar light source at 60 s, 180 s, and 300 s respectively; Temperature changes at **(k)** tail and **(l)** head of G-TPU layer arrow shaped structure with [0°/0°], [0°/90°] and [90°/90°] stacking manners.

As is well known, the IP thermal conductivity of graphene is as high as 3000 W/(m·K),

while its thermal conductivity in the thickness direction is less than 5 W/(m·K).[16] Therefore, when graphene is oriented along the printing direction, its high IP thermal conductivity can significantly enhance the thermal conductivity in that direction. The thermal conductivity of samples with different stacking manners exhibited similar values in the TP direction. This consistency indicated that the stacking manner had minimal impact on the thermal conductivity in the TP orientation of the samples. But in the IP direction the thermal conductivity shows trend of $TC_{\parallel \text{ Filling}} > TC_{+ \text{ Filling}} > TC_{\perp \text{ Filling}}$. As previously noted, a key distinction of FDM printing was its layer-by-layer stacking process, which lead to distinct morphologies at the interlayer interfaces. Although the nozzle used for extruding the molten material was circular (diameter = 0.4 mm), the material was compressed during deposition from spherical beads to a thickness of 0.1 mm, becoming elliptical and consequently creating gaps (voids) between adjacent deposited filaments,[43–45] as confirmed in **Figure 2b**. These voids function as thermally insulated regions that detrimentally affected thermal conductivity. Additionally, the presence of voids diminished the connection strength between adjacent filaments. However, the extent of this impact varied with the distribution of the voids. To achieve high thermal conductivity, a continuous heat conductive pathway, incorporating a network of thermally conductive materials, was essential in the IP direction, and the orientation of the filler should align with this direction. These two factors were critical in influencing thermal conductivity in practical applications.[33]

To directly observe the thermal conduction among different samples, an arrow shaped G-TPU single-layer structure was printed with different stacking manners (with dimensions described in **Figure S6**). Shown in **Figure 4a,** a converged simulated solar light source with an intensity of 0.4 W/cm² was used to illuminate the tail of arrow samples printed in three distinct stacking manners for 300 s. Infrared images and temperature changes at the head of these samples were systematically monitored and recorded using an infrared camera. **Figure 4b** and **4c** summarized the temperature variations at the tail and the head of the arrow sample over time with exposure to light and the recording time interval was 30 s, respectively. Due to the superior photothermal conversion properties of graphene, the temperature at the tail end of the arrow sample rapidly increased within the first 30 s of light exposure. Concurrently, the temperature at the arrow's head slowly increased as the generated heat from tail was transferred to the head of the arrow. The representative infrared images of arrow samples printed in a [0°/90°] stacking manner subjected to irradiation from a converged simulated solar light source at intervals of 60 s, 180 s, and 300 s were shown in **Figure 4d – 4f**, respectively, and other infrared images of characterization were shown in **Figure S7**. The positions of the arrow samples were clearly marked in the figure. An intuitive observation from these images shows that as the exposure time increases, the rainbow aperture at the tail of the arrow gradually expands, with the outermost aperture extending progressively toward the head of the arrow. The temperature at the tail end of all samples generally reached equilibrium after 180 seconds. Notably, the temperature changes at the tail end of the samples stacked by [90°/90°] were the fastest over time, with the final equilibrium temperature reaching a maximum of 103 °C. The final equilibrium temperatures of the samples stacked by [0°/90°] and [0°/0°] were comparable and both lower than those stacked by [90°/90°]. However, the heating rate of the [0°/90°] stacked samples was higher than that of the [0°/0°] stacked samples. Conversely, the temperature at the head of the arrow samples exhibited an opposite trend. Specifically, the head of the arrow sample printed in the [0°/0°] stacking manner heated the fastest within 300 s, reaching the highest final temperature of 46 °C. The final equilibrium temperature of the [0°/90°] stacked samples was similar to that of the [0°/0°] stacked samples, but with a slower heating rate. The [90°/90°] stacked arrow samples exhibited the slowest heating rate over the same period, with a final temperature of only about 39 °C.

In short, the variance in heating rates and final temperatures between the tail and the head of the arrow samples printed by the three stacking methods underscored the differences in thermal conductivity. In particular, the [90°/90°] stacked samples demonstrated the lowest IP thermal conductivity, resulting in heat accumulation at the tail end and inefficient heat conduction; thus, the tail end heated up quickly and reached a higher final temperature, while the head remained cooler. In contrast, the [0°/0°] stacked samples had a smooth and continuous thermal conduction path in the IP direction, allowing heat to be efficiently conducted to the top end. Consequently, the tail end of these samples registered the lowest temperature while the top end reached the highest temperature. The temperature variation between the tail and head of the [0°/90°] stacked sample lied between those of the previously mentioned two stacking manners. A rational design of the printing path could optimize the anisotropic thermal conductivity of 3D-printed microfilaments. Owing to the layer-by-layer processing inherent to the FDM method, the printing direction of each layer can be digitally adjusted. This adjustability provided control over the thermal conductivity of the printed parts.[46] In practical applications, this allowed for customization according to specific requirements, highlighting the broad potential for diverse applications.

Furthermore, as described in **Figure 3a – 3c**, although the orientation of graphene flakes within samples using three different stacking manners were similar in the IP direction, the abundant voids in products fabricated with the [90°/90°] stacking manner obstructed the formation of heat conductive pathways in the IP direction, thereby impeding heat flux transfer. Conversely, in samples with the [0°/0°] stacking manner, heat was efficiently transferred along continuous filaments aligned in the same direction, specifically, the IP direction. The fluent heat conductive pathways uninterruptedly exist with the voids parallelly aligned. This configuration allowed thermal energy to transfer smoothly along the continuous heat conductive pathways. However, for samples utilizing the [0°/90°] stacking manner, the weak adhesion between adjacent filaments relying solely on surface contact that disrupted the heat transfer pathways in half of the total layers due to the "cross" filling method and the weak adhesion resulted in increased thermal resistance.[47] In contrast, heat transfer in neat polymers primarily relied on lattice vibrations, specifically phonon vibrations, since free electrons were scarce. Consequently, the orientation of polymer molecular chains along the nozzle scanning direction contributed to a more regular chain structure. This alignment enhanced thermal conductivity to a certain extent.[48] Hence, with consistent graphene orientation, the trend in thermal conductivity was depicted in **Figure 3h**. Eliminating all voids was challenging due to the inherent working principles of FDM printing. However, larger voids can be minimized into smaller ones by optimizing FDM parameters, such as layer height, printing speed, and others.[49]

To illustrate the anisotropic heat transfer mechanism of G-TPU layer arrow shaped structure, we performed finite element numerical simulations. Shown in **Figure 4g**, the arrow shaped structure was modeled where the pink region indicated the illumination region of the converged light at tail. The heat transfer module with coupled interfaces of radiative beam in absorbing media was applied. The parameters of the materials (listed in **Table S2**) were applied, and the boundary conditions of convective heat flux (upper layer and side walls of the structure), thermal insulation (bottom layer of the structure) and illumination region with input power of 0.4 W/cm² in the pink region were set. The overall structure was finely meshed and then the computation was performed with 1,000,000 degrees of freedom.

**Figure 4h – 4j** showed the representative calculated temperature distribution of G-TPU/N-TPU double-layer arrow samples printed in [0°/90°] stacking manners, irradiated by a simulated solar light source at 60 s, 180 s, and 300 s respectively. These heat maps indicated that the accumulated heat in the arrow tail region would transport along the longitudinal axis of the arrow, and eventually, raise the arrowhead temperature. By taking anisotropic thermal conductivity of G-TPU layer into the simulation, and the averaged temperature at arrow head and arrow tail of different stacking manners were evaluated and summarized in **Figure 4k** and **4l**, respectively. In consistent to the trends in the experimental observations mentioned above, both the arrow head temperature and arrow tail temperature raised as the illumination at arrow tail. To specific, the [90°/90°] stacked samples showed highest equilibrium temperature of 109 °C at the tail, followed by 101 °C and 93 °C for [0°/90°] and [0°/0°], respectively, which was in excellent agreement to the experimental value of 103 °C, 95 °C and 92 °C. While their corresponding temperature at the head were 39 °C, 45 °C, and 52 °C, comparing to the experimental measurement of 39 °C, 42 °C, and 46 °C.

This simulation results together with our experimental observations proved that the highly aligned graphene flakes from extruded G-TPU filament contributed the to the enhancement of directional thermal conductivity.

## 3. Heat Retention Effect

It was anticipated that the G-TPU/N-TPU double-layer structure would have heat retention effect as its low thermal conductivity in the TP direction. To prove this, the single-layer structure of G-TPU (with a thickness of 0.5 mm) to G-TPU/N-TPU double-layer structure (with an overall thickness of 1 mm) were assembled into an array (**Figure 5a**) for comparison. The printed single-layer G-TPU structure and G-TPU/N-TPU double-layer structure (indicated in red arrow) was identical in shape and color. The matrix was baked in oven for 10 minutes to reach the temperature of ~68 °C as observed though an IR camera in **Figure 5b.** Subsequently, the matrix was transferred to ambient environment, and the temperature distribution of the matrix was shown in **Figure 5c** that the G-TPU/N-TPU double-layer structure showed higher temperature than G-TPU single-layer structure after 30 s, which indicated the excellent heat retention of G-TPU/N-TPU double-layer structure.

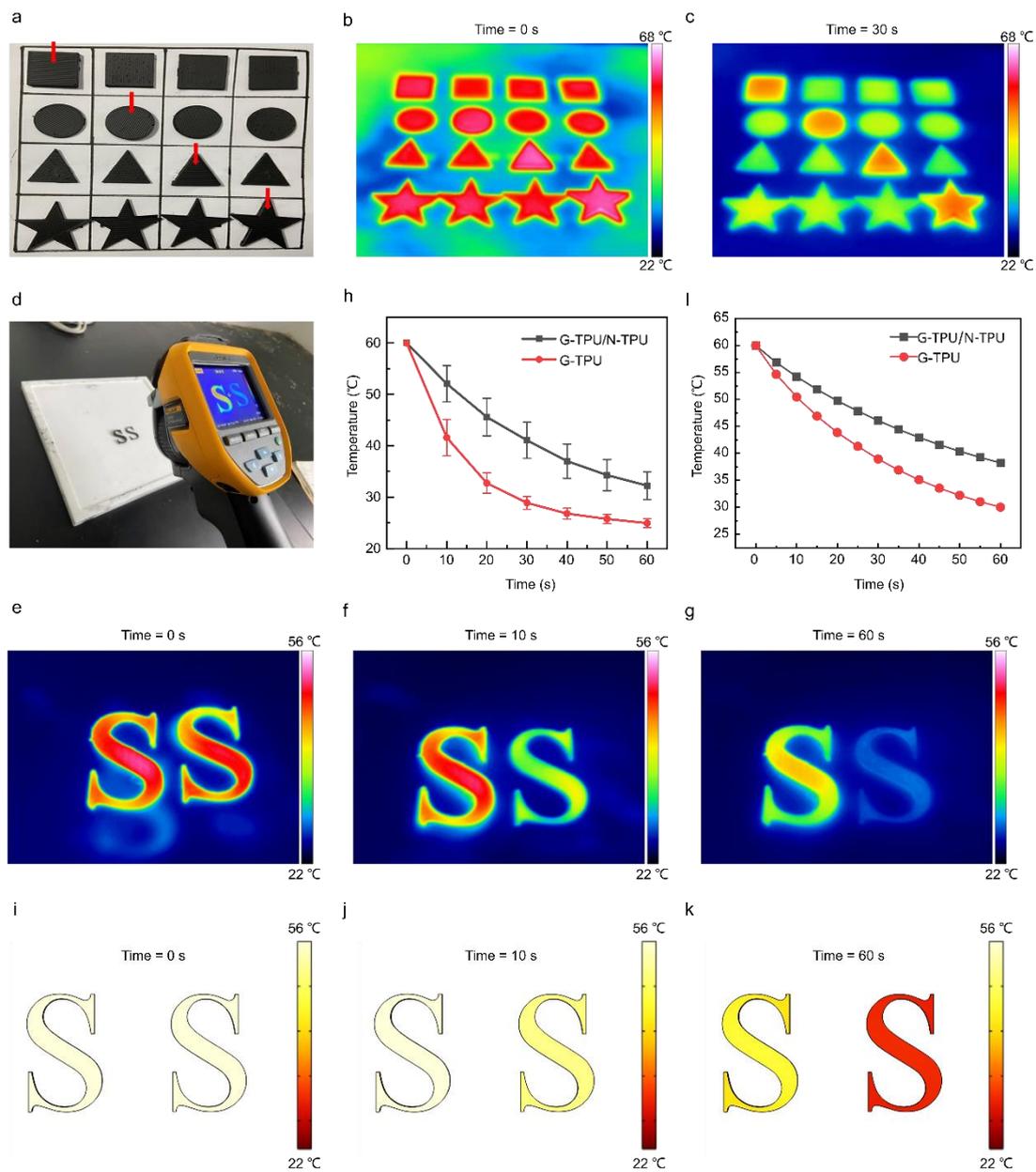

**Figure 5** Study on heat retention effect of G-TPU/N-TPU double layer structures. **(a)** 3D printed array containing G-TPU/N-TPU double layer and single-layer G-TPU structure, the red arrow indicated G-TPU/N-TPU double layer structures; **(b)** Infrared image of the baked array during the heat retention test at 0 s; **(c)** Infrared image of the baked array during the heat retention test at 30 s; **(d)** Heat retention test setup of printed letters; **(e) – (g)** Infrared image of the printed letters during the heat retention test at 0 s, 10 s and 60 s, respectively; **(h)** Temperature changes of printed letters during the heat retention test as function of time; **(i) – (k)** Calculated heat distribution of the printed letters during the heat retention test at 0 s, 10 s and 60 s, respectively; **(l)** Calculated temperature changes of printed letters during the heat retention test as function of time.

To systematically study this heat retention effect, as displayed in (**Figure 5d** and **Figure S8)**, a G-TPU/N-TPU double-layer sample of letter "S" with a thickness of 1 mm (left) and a G-TPU single-layer sample of letter "S" (right) with a thickness of 0.5 mm were printed, respectively. Both samples were baked in oven for 10 minutes to reach equilibrium temperature of ~60 °C and were subsequently transferred to ambient environment, and the temperature distribution of the samples were monitored and recorded via an IR camera. The representative experimental measured temperature distributions of printed letters during the cooling were shown in **Figure 5e – 5g**, and the temperature changes of each letter were summarized in **Figure 5h**. Initially, G-TPU/N-TPU double-layer structure exhibited roughly the same temperature as single-layer G-TPU structure. However, the single-layer G-TPU structure dropped to 34 °C within 20 s and becoming barely visible on the infrared camera by 60 seconds. In stark contrast, the temperature of the G-TPU/N-TPU double-layer sample remained at 35 °C even at 60 s. This can be attributed to the fact that the bottom layer of N-TPU in the double-layer structure serves as a thermal insulation layer in the TP direction that further slowed the heat loss.

To further validate the heat retention effect of G-TPU/C-TPU double-layer structure, numerical simulations were performed. Shown in **Figure 5i – 5k**, two "S" shaped structures of double-layer (left) and single-layer (right) were developed from 3D printing models. Calculation was performed with experimental measured parameters (listed in **Table S2**) and defined boundary conditions (thermal insulated bottom layers and natural convective cooling heat flux to the external environment with upper layers and walls). The calculated temperature changes of each letter were summarized in **Figure 5l**, in consistent to the experimental measured values that G-TPU/N-TPU double-layer structure exhibited better heat retention performance than single-layer G-TPU structure.

Above results and discussion clearly demonstrated that the double-layer structure offered superior heat retention properties. This performance, in conjunction with the previously discussed excellent thermal conductivity in printing direction, underscored the anisotropic thermal conductivity of G-TPU/N-TPU double-layer structure.

## 4. Photothermal Performance

The demonstrated anisotropic thermal conductivity of G-TPU/N-TPU double-layer structure indicated the superior potential of the structure for heat collection. The structure's performance in photothermal conversion was then tested.

First, the G-TPU/N-TPU double-layer structure was evaluated through a de-icing test. Specifically, the photothermal de-icing performance of C-TPU (purchased commercial black colored filament) single-layer structure sample with 1 mm thickness and a G-TPU/N-TPU double-layer structure sample of the same thickness were compared. As illustrated in **Figure 6a**, 3 mL of water was placed on the printed sample surface, and the samples were then stored in a freezer at -83 °C until the water droplets frozen into ice. Subsequently, the two samples were transferred and positioned on a 3D-printed inclined plane set at a 30-degree angle relative to the horizontal, with the ambient temperature maintained at 0 °C. A simulated solar light source with an intensity of 0.15 W/cm² was used to irradiate the sample surfaces, and the state of the ice particles on the surface was observed and recorded with a camera. As shown in **Figure 6b**, the G-TPU/N-TPU double-layer structure sample had melted half of the ice volume after 83 seconds, with the ice gradually sliding off as it decreased in size. After 192 seconds, the ice particles were completely melted, leaving only water residues adhered to the sample surface. In contrast, the sample on single-layer structure was only melted for approximately one-third of the ice volume after 124 seconds and was not fully melted even after 259 seconds. As depicted in **Figure 6c,** when exposed to sunlight, the G-TPU layer exploited graphene's strong absorption characteristics across the wide-band solar spectrum, coupled with a large absorption surface and efficient photothermal conversion capabilities.[50] This configuration facilitates rapid heat collection and heating effects. Due to graphene's high anisotropic thermal conductivity, heat flow was quickly conducted and evenly distributed within the G-TPU layer, leveraging graphene's superior in-plane thermal conductivity. Additionally, the low thermal conductivity of the TPU material in the bottom layer of the structure and the limited through-plane thermal conductivity of graphene impeded heat loss in the TP direction, thereby enhancing the heat retention properties of the double-layer structure. As demonstrated above, the G-TPU/N-TPU double-layer structure exhibited excellent photothermal de-icing performance.

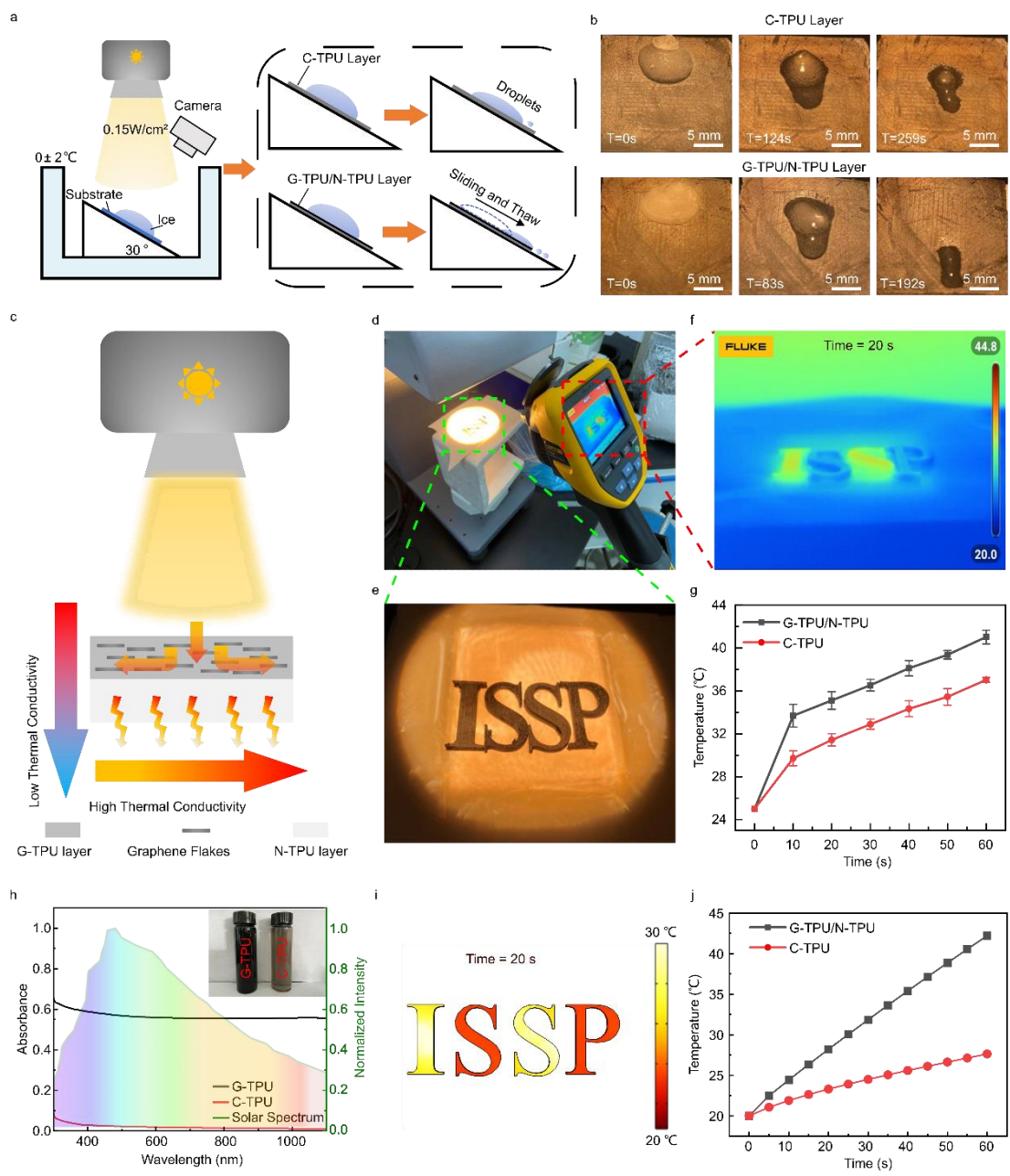

**Figure 6** Study on photothermal performance of G-TPU/N-TPU double-layer structures. **(a)** Schematic diagram illustrated the experimental setup for photothermal de-icing test; **(b)** Photographs showed the de-icing performance of G-TPU/N-TPU double-layer structure and C-TPU single-layer structure; **(c)** Schematic diagram illustrated the heat flow in G-TPU/N-TPU double layer structures during the photothermal conversion; **(d)** Experimental setup showed the photothermal-IR imaging test; **(e)** 3D printed samples for photothermal-IR imaging test; **(f)** Infrared image of the sample after 20 s exposure to a simulated solar light source; **(g)** Experimental measured temperature changes of the illuminated G-TPU/N-TPU double-layer structure and C-TPU single-layer structure as function of time; **(h)** UV-Vis-NIR spectrum of G-TPU filament and C-TPU filament solutions (2 mg/mL in DMF) and normalized intensity of simulated solar light source; **(i)** Calculated temperature distribution of the test samples after 20 s illumination; **(j)** Calculated temperature changes of the illuminated G-TPU/N-TPU double-layer structure and C-TPU single-layer structure as function of time.

Since the huge difference of the surface temperature of these two structures, their potential for infrared anti-counterfeiting applications were further evaluated. As depicted in **Figure 6d**, photothermal experiments were conducted under a solar simulator, with a solar power meter measuring the intensity of simulated sunlight at 0.15 W/cm², and a portable infrared thermal camera was employed to capture infrared images and record temperature fluctuations. The prepared samples were shown in **Figure 6e** (**Figure S9** illustrated the representative design model), that the letters "I" and "S" were printed as G-TPU/N-TPU double-layer structure, while letters "S" and "P" were printed with C-TPU filament to form the acronym "ISSP" (Institute of Solid State Physics). It is virtually impossible for the naked eye to differentiate between samples printed with the two different filaments. This observation is further corroborated by the changes in the samples' transmittance across the 400 – 2000 nm wavelength range (**Figure S10**). However, as shown by the IR image in **Figure 6f**, it is evident that after 20 seconds of sunlight exposure, samples printed with the letters "I" and "S" of double-layer structures were more distinctly visible on the infrared camera compared to those single-layer structure printed solely with C-TPU. As illustrated in **Figure S11**, it can be observed that within a 60 seconds illumination period, the colors captured by infrared cameras shifted from cool to warm tones. Notably, the infrared images of the double-layer structure samples were consistently more pronounced than those of the C-TPU samples. To accurately determine the temperature of the printed samples, the recorded infrared images were analyzed and quantified using SmartView Classic 4.4 software, with the temperature variations over time documented in **Figure 6g**. The double-layer structure samples exhibited temperatures ~ 4 °C higher than those of the single-layer C-TPU samples throughout the testing process. Specifically, at 10 seconds of light exposure, the temperature of the double-layer structure samples reached 34 °C, while the C-TPU samples reached only 29.5 °C. Smaller-sized graphene sheets generally exhibited enhanced dispersion stability, facilitating the formation of a denser and more uniformly dispersed structure within the composite matrix. This characteristic enabled more consistent heat generation upon irradiation. Additionally, other studies reported that smaller 2D materials were more beneficial in avoiding void formation than the large-sized ones.[51] The huge contrast in infrared images and the differential in temperature changes effectively demonstrated the superior heat collection and heating effect of the designed double-layer structure and hold significant potential for applications in infrared anti-counterfeiting.

To further validate the photothermal performance of these two structures, numerical simulations were performed. Similar to the simulation of G-TPU/N-TPU double-layer structure's heat retention effect mentioned above, letters of "ISSP" were modeled based on

3D printing models. To quantify the optical density of G-TPU and C-TPU layers, as shown in the **Figure 6h** and its inset, the UV-Vis-NIR spectrum of G-TPU and C-TPU filament solution (2 mg/mL in DMF) were measured. It can be seen from the inset photograph that the G-TPU solution was much darker than that of C-TPU, and the spectrum of their dissolved solution in DMF clearly indicated that the G-TPU layer containing graphene flakes had higher optical density than C-TPU layer containing carbon black at nearly all wavelength of interest. Based on the estimated absorption coefficients of G-TPU and C-TPU from **Table S2**, the calculated heat map of the structure in **Figure 6i** matched the experimental characterized IR image in **Figure 6f**. And their estimated temperature changes in **Figure 6j** generally followed the same trend as the observation in the experiments.

## 5. Mechanical Property

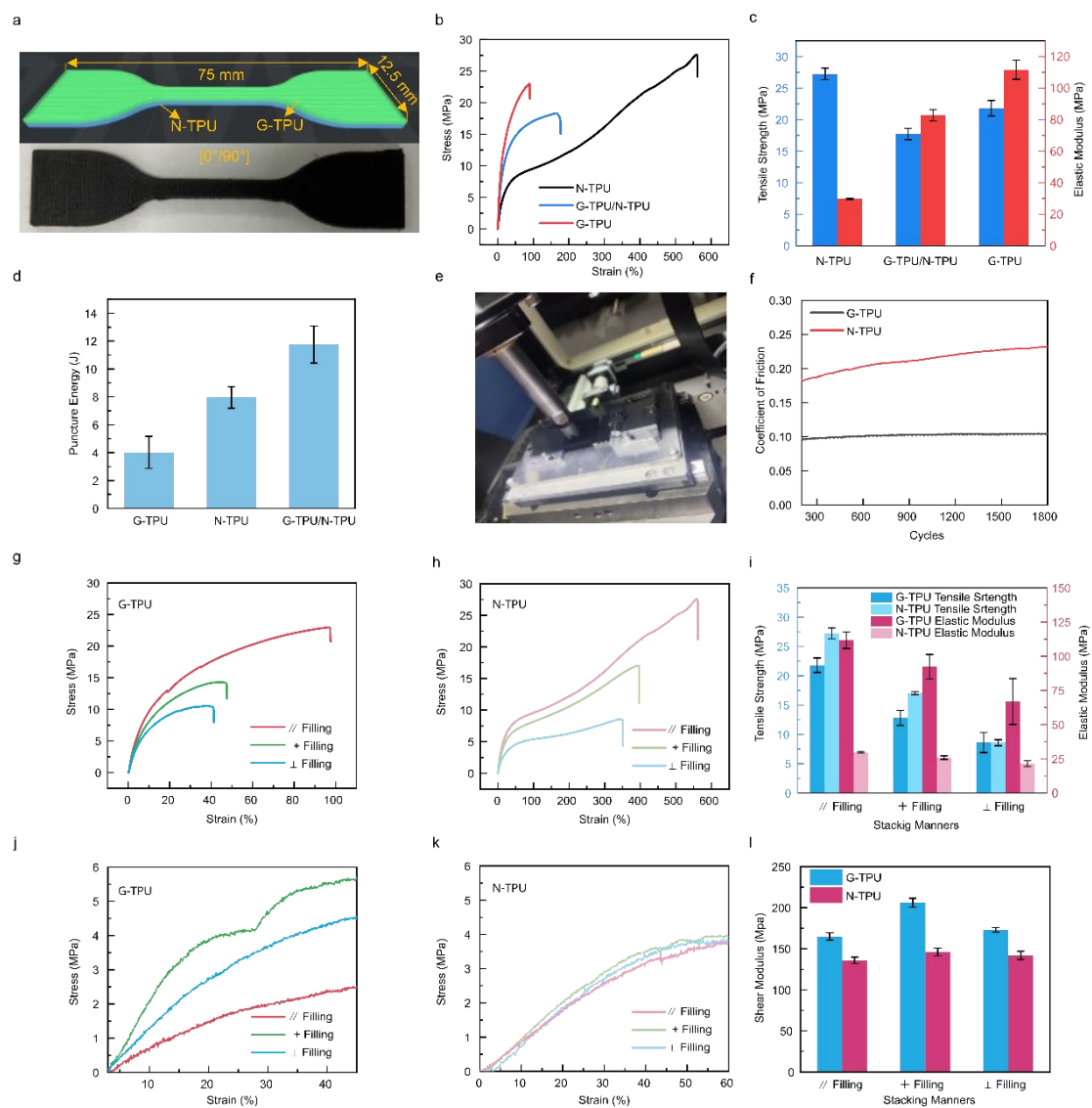

**Figure 7** Experimental characterization of G-TPU/N-TPU double-layer structures. **(a)** Design model and printing of the G-TPU/N-TPU double-layer sample model stacked in [0°/90°] cross filling; **(b)** Stress-strain curves of [0°/0°] stacked samples for N-TPU, 30 wt% G-TPU single-layer structures, and 30 wt% G-TPU/N-TPU double-layer structures; **(c)** Tensile strength and elastic modulus of N-TPU, 30 wt% G-TPU single-layer structures, and 30

wt% G-TPU/N-TPU double-layer structures; **(d)** Puncture energy of N-TPU, 30 wt% G-TPU single-layer structures, and 30 wt% G-TPU/N-TPU double-layer structures; **(e)** Scheme of the friction testing machine; **(f)** Coefficient of friction for N-TPU and 30 wt% G-TPU samples over 1800 cycles; Stress-strain curves of **(g)** 30 wt% G-TPU and **(h)** N-TPU in three different stacking manners; **(i)** Tensile strength and modulus for 30 wt% G-TPU and N-TPU in three different stacking manners; Shear stress-strain curves of **(j)** 30 wt% G-TPU and **(k)** N-TPU in three different stacking manners; and **(l)** Shear modulus of 30 wt% G-TPU and N-TPU in three different stacking manners.

The mechanical properties of 3D-printed objects were crucial for practical applications. First, the elastic modulus of these printed objects were studied. Shown in **Figure 7a**, the G-TPU/N-TPU double-layer structure sample with [0°/90°] stacking for mechanical performance testing was modeled via CAD software and processed in slicing software prior to 3D printing (**Figure 7a upper image**), and alongside the actual printed sample (**Figure 7a lower image**). The printed double-layer structure was compared to N-TPU and G-TPU single-layer structure with the same stacking manners in terms of elastic modulus test. As shown in **Figure 7b** and **Figure 7c**, the graphene-enhanced samples showed higher elastic modulus than that of neat TPU, while neat TPU exhibited the highest elongation at break. This was because, in composites, the greater freedom of molecular chain movement leads to higher elongation at break. Graphene flakes, as fillers, adsorbed and entangled polymer chains, restricting their movement and thereby reducing the elongation at break.[52] Meanwhile, compared to the G-TPU single-layer structure, the G-TPU/N-TPU double-layer structure exhibited a twofold increase in elongation at break showing excellent mechanical property.

Additionally, as shown in **Figure 7d**, impact tests were conducted on these three types of samples. The G-TPU/N-TPU double-layer structure demonstrated excellent impact resistance, with a puncture energy reaching 12.86 J. The high elastic modulus of the G-TPU layer provides strong resistance during the initial impact stage, while the softer N-TPU layer absorbs and dissipates energy in the subsequent stages. This combination effectively minimized the transmission of puncture force. The lubricating characteristics of the coating were analyzed using a reciprocating ball-on-disk tribometer (**Figure 7e**). **Figure 7f** showed the friction coefficient curves of N-TPU and G-TPU surfaces. The coefficient of friction of the G-TPU sample surface was lower and more stable (COF ~0.1) compared to N-TPU, whose friction coefficient increased with the number of friction cycles. Notably, the low COF of the G-TPU surface could withstand approximately 1800 friction cycles, indicating excellent wear resistance. Despite high strength, the introduced graphene flakes enchanted the composite with excellent lubrication. When graphene flakes were uniformly dispersed in the TPU matrix, the lubricating effect of graphene could reduce the direct contact area of friction surfaces, thereby lowering the coefficient of friction.

Moreover, the enhancement of graphene flakes in TPU under different stacking manners was also investigated. The stress-strain curves of 30 wt% G-TPU and N-TPU samples with different stacking manners were shown in **Figure 7g** and **Figure 7h**, and their tensile strength and elastic modulus were compared in **Figure 7i**. The introduction of graphene flakes led to the enhancement of elastic modulus and decreased elongation at break with all different stacking manners. Among these, the [0°/0°] samples exhibited the highest strain, tensile strength, and elastic modulus. This was attributed to the alignment of the printed filaments with the direction of the applied force as well as the strong interlayer bonding, highlighting the importance of stacking design. **Figure 7j** and **Figure 7k** showed the shear stress-strain curves for 30 wt% G-TPU and N-TPU in different stacking manners, respectively, and **Figure l** presented their shear modulus comparison. It could be observed that the high strength and modulus of graphene enhanced the overall shear performance of the G-TPU composite. Additionally, the [0°/90°] stacked samples, printed with filaments alternately stacked in different directions, demonstrated the highest shear strength and modulus due to the uniform stress distribution. N-TPU samples exhibited similar trends in stress-strain and shear modulus with different stacking manners as seen in the G-TPU samples. However, the differences between the various stacking manners are less pronounced for N-TPU, likely due to its isotropic nature.

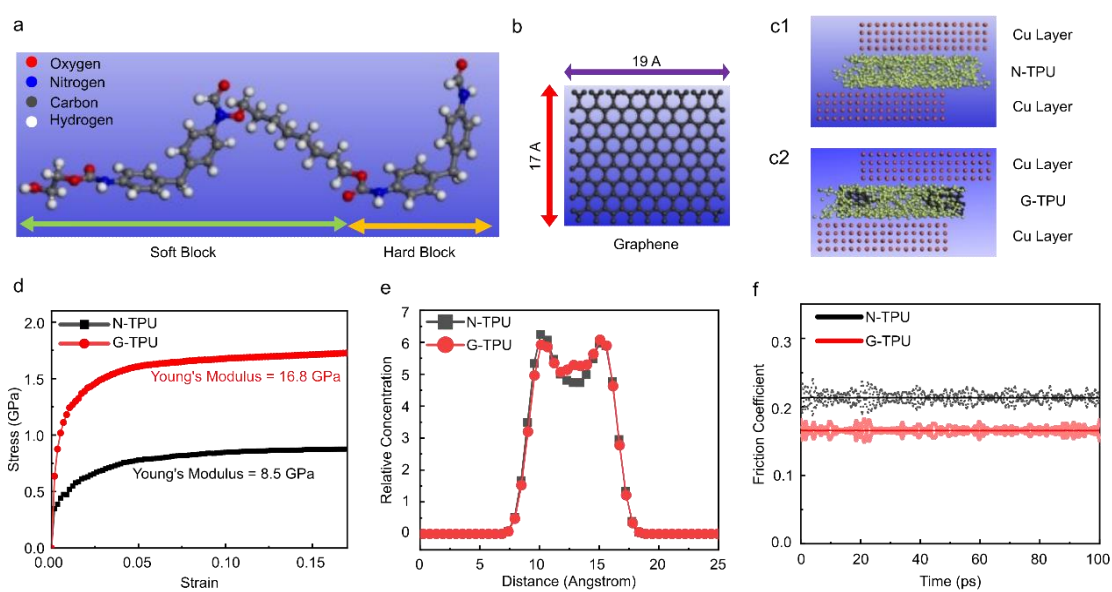

**Figure 8** Molecular simulation of the mechanical properties of G-TPU and N-TPU. **(a)** Modeled TPU repeating unit; **(b)** Modeled graphene flake; Modeled N-TPU layer **(c1)** and G-TPU layer **(c2)** for confined shear simulation; **(d)** Calculated stress-strain curves and estimated Young's Modulus of N-TPU and G-TPU; **(e)** Calculated relative atoms concentration of N-TPU and G-TPU as the height layer in **Figure 8c1** and **8c2**; **(f)** Calculated friction coefficient of N-TPU and G-TPU.

To further study the enhancement of graphene flakes in the polymer composite at atomic and molecular level, molecular dynamics simulation was performed with Materials Studio software. Shown in **Figure 8a**, the repeating unit of TPU was modeled with molar ratio of soft block to hard block as 1:1 after geometric optimization with Dmol3 module. The graphene flake was described as monolayer with dimension of 17 A × 19 A in **Figure 8b**. The N-TPU and G-TPU (30 wt% graphene) box were built with Amorphous Cell module, and subsequently assembled into layers for mechanical properties study and confined shear simulation via Forcite tools as **Figure 8c1** and **8c2,** where the green regions indicated the TPU polymer chains and the dark regions suggested the graphene flakes. The calculated stress-strain curves N-TPU and G-TPU was shown in **Figure 8d**, which clearly proven that G-TPU had much higher Young's Modulus (16.8 GPa) than N-TPU (8.5 GPa). It should be noticed that compared to the aforementioned experimental characterization in **Figure 7g – 7l**, the calculated values were higher due to the inevitable structural flaws and defects in the printed samples. **Figure 8e** summarized calculated relative atoms concentration of N-TPU and G-TPU as the height layer. It was interesting that the atom concentration in G-TPU (red dots and lines) was higher than the N-TPU at the center region of the sandwiched layer structure, which indicated the intense interaction between the polymer chains and graphene flakes. Shown in **Figure 8f**, the friction coefficients of N-TPU and G-TPU during the confined shear simulation were further evaluated, and G-TPU (~0.22) showed weaker friction interaction than N-TPU (~0.17).

Based on the systematical study mentioned above, it could be proved that the introduction of graphene flakes improved the overall mechanical properties of the composite. The improvements in terms of modulus and lubrication would guarantee the photothermal application of G-TPU/N-TPU double-layer structure at vast.

## Conclusion

In this study, a novel G-TPU/N-TPU double-layer structure was developed via independently printing graphene-thermoplastic polyurethanes (G-TPU) and neat thermoplastic polyurethanes (N-TPU) with a double-nozzle FDM 3D printer. The double-layer structure exhibited anisotropic thermal conductivity with aspect ratio of ~8 as the introduced graphene flakes enhanced thermal conductivity in the G-TPU layer and thermal insulation nature of N-TPU layer. Then its high IP thermal conductivity was evaluated, which was further affected by stacking manners due to the directional alignment of graphene flakes during the printing process. The heat retention effects due to the low TP thermal conductivity was also studied. Due to these unique thermal characteristics, combined with extraordinary photothermal conversion rate of graphene, the G-TPU/N-TPU double-layer structure showed superior photothermal performance in de-icing test and photothermal IR imaging test, which showed ~4 °C higher than that of the colored black TPU (C-TPU) material and further confirmed with finite element method (FEM) simulations. Nonetheless, the imported graphene flake's enhancement in mechanical properties was also investigated. Impact test results (puncture energy of 12.86 J) together with friction test results (coefficient of 0.1 over 1000 cycles) demonstrated excellent impact and wear resistance that suitable for photothermal applications in severe environment. Molecular dynamic (MD) simulations results depicted that the enhancement was induced by the strong interaction between polymer chains and graphene flakes. It was envisioned that the developed G-TPU/N-TPU double layer structure with anisotropic thermal conductivity was vast field of photothermal applications.

## Experimental Section

Essential Experimental Procedures/Data are available in the Supporting Information of this article.

## Supporting Information

Supporting Information is available from the Wiley Online Library or from the author.

# Acknowledgement

Mr. Zihao Kang performed the experiments and draft the manuscript. Dr. Min Xi designed the experiment, performed the simulations and amended the manuscript. The support and supervision from Prof. Nian Li, Prof. Shudong Zhang and Prof. Zhenyang Wang during the research were appreciated.

This work was financially supported by the National Key Research and Development Project (2022YFA1203600, 2020YFA0210703), the National Natural Science Foundation of China (Nos. 12204488).

# Conflict of Interest

The authors declare no conflict of interest.

# Supporting Information

# Anisotropic Thermal Conductivity of 3D Printed Graphene Enhanced Thermoplastic Polyurethanes Structure toward Photothermal Conversion

Zihao Kang[a], Min Xi*[b], Nian Li[b], Shudong Zhang[b], Zhenyang Wang*[b]

[a] School of Energy, Materials and Chemical Engineering, Hefei University, Hefei, 230601, P. R. China

[b] Institute of Solid State Physics and Key Laboratory of Photovoltaic and Energy Conservation Materials, Hefei Institutes of Physical Science, Chinese Academy of Sciences, Hefei, Anhui 230031, P. R. China

Corresponding authors:

Min Xi − orcid.org/0000-0003-4414-3110; Email: minxi@issp.ac.cn

Zhenyang Wang − orcid.org/0000-0002-0194-3786; Email: zywang@iim.ac.cn

# Experimental Methods

## Materials

Graphene flakes (carbon content ~95%) with a lateral size of ~30 μm were bought from Knano Graphene Technology Co., Ltd. The morphologies of graphene flakes are shown in **Figure S1**. Thermoplastic polyurethane (TPU) with hardness of 95A was purchased from Toprene Enterprise Co., Ltd. Colored TPU (black) was purchased from Raised 3D technology Co., Ltd. N, N-dimethylformamide (DMF, AR, 96%) was provided by Aladdin Industrial Co., Ltd. (Shanghai, China). All chemicals were used without further purification.

## Preparation of G-TPU Filaments.

The fabrication of G-TPU filament was based on a two-step mixing procedure (solution mixing and melt blending) which has been reported before.[1] Specifically, the TPU pellets were first dissolved in DMF solution under though mechanical stirring for 12 h at room temperature. Subsequently, graphene flakes of different ratios (0 wt%; 5 wt%; 10 wt%; 20 wt%; 30 wt%) were added into the solution and subjected to ultrasonic treatment to ensure uniform dispersion, followed by continuous stirring. The resulting viscous solution was then dried in an oven at 100 °C for 15 h to remove excess solvent, and then a black film could be obtained. The G-TPU film was manually cut into small pieces and melted by WLG10G twin-screw extruder (Shanghai Xinshuo Precision Machinery Co., Ltd., China). Upon cooling and cutting, a composite masterbatch was produced. The fabricated masterbatch was extruded into filament using a D12 single screw extruder (Shanghai Xinshuo Precision Machinery Co. Ltd., China). The extrusion process was meticulously controlled by setting the screw speed and temperature at 30 rpm and 200 °C, respectively, and a standard composite filament with diameter of 1.75 mm was produced for FDM printing.

## Preparation of 3D Printed Samples.

All filaments were dried at 60 °C for 5 hours before printing. The printing was performed using a dual-nozzle 3D printer (Raise 3D Pro3) equipped with a nozzle diameter of 0.4 mm. All printing parameters were established and controlled via ideaMaker software and were listed in **Table S1**. The term "stacking manners" referred to the angle between the melt filling direction and the printing direction. For convenience, the double-layer structural sample, printed with G-TPU filament and neat TPU filament, was abbreviated as G-TPU/N-TPU. Three stacking manners ([0°/0°], [0°/90°], [90°/90°]) were referred to as ∥ filling, + filling, and ⊥ filling, respectively. All samples were produced using a dual-nozzle 3D printer (Raise 3D Pro3) equipped with a nozzle diameter of 0.4 mm. The 3D models were designed using Solidworks software and sliced by ideaMaker software. Within ideaMaker software, various printing paths and stacking manners were configured, and the temperature was maintained at approximately 230 °C with a printing speed of 30 mm/s, and the printing bed was set at 60 °C. The layer thickness was set to 0.1 mm and the filling density to 100 % to ensure a dense and finely structured composition for all samples.

Table S1 The main printing parameters of FDM

| Printing parameters | Value |
| --- | --- |
| Nozzle temperature (°C) | 230 |
| Platform temperature (°C) | 60 |
| Left nozzle diameter (mm) | 0.4 |
| Right nozzle diameter (mm) | 0.4 |
| Layer height (mm) | 0.1 |
| Filling rate (%) | 100 |
| Printing speed (mm/s) | 30 |
| Flow compensation | 1.2 |
| Stacking manner (°) | [0°/0°][0°/90°][90°/90°] |

**Morphological Characterizations.**

The surface morphology of graphene flakes was observed with FESEM (SU8020, HITACHI, Japan) with the acceleration voltage of 10 kV. As shown in **Figure S1**, the samples were deposited on a silicon wafer and gold coated before characterization.

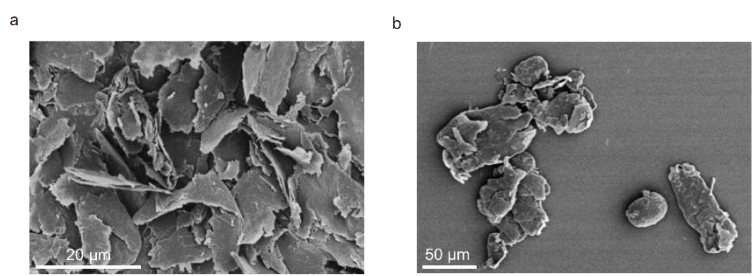

**Figure S1** SEM images of graphene flakes.

The surface morphology of filament was obtained by FESEM (SU8020, HITACHI, Japan) with the acceleration voltage of 10 kV. The samples were immersed in liquid nitrogen for 15 minutes and then brittlely fractured. After gold coating treatment, the morphologies were observed.

The composites masterbatch samples were dispersed in resin after sonication into short nanofibers and then were cut into thin slices of thickness of ~100 nm with an ultrathin nano-blade slicer. And the sliced samples were characterized by transmission electron microscopy (TEM, JEM-2100F, Japan) at the accelerating voltage of 160 kV. The morphologies of the samples were presented in **Figure S2**.

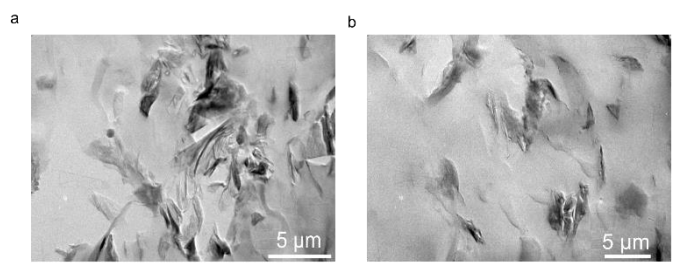

**Figure S2** TEM images of graphene TPU composites masterbatch sample with 30 wt% graphene content.

**Thermogravimetric Analysis (TGA).**

Thermogravimetric analysis (TGA2, METTLER TOLEDO, Switzerland) to assess the composition and thermal stability of various composites were shown in **Figure S3**. Approximately 10 mg of each sample was heated from room temperature to 800 °C at the rate of 10 °C/min under nitrogen atmosphere. The residual weights of the different samples were subsequently measured.

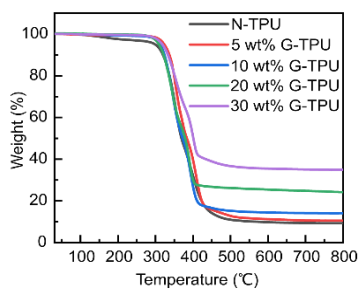

**Figure S3** TGA curves of G-TPU filaments with different graphene contents.

**Raman Spectroscopy Characterization.**

Raman spectroscopy (DXR3, ThermoFisher, America) equipped with a He-Ne laser excited at 532 nm was employed to characterize the quality of the graphene flakes.

**Rheological Tests.**

The rheological properties were tested using a rheometer (HR-20, TA Instruments, USA). Measurements were performed over a frequency range of 0.1 to 100 rad/s with a strain amplitude of 1 % at 230 °C.

**XRD and WAXS Characterization.**

An X-ray diffractometer (SmartLab SE, Rigaku, Japan) utilizing Cu Kα radiation ($\lambda$ = 0.15406 nm) was used to characterize the crystal plane diffraction peaks of graphene to determine the orientation of graphene flakes in different printed products. The scan angle range of 10 to 80° at a scan speed of 10 °/min. Voltage and current intensity are 40 kV, 100 mA respectively. Samples prepared for XRD testing were 3D printed at a 20 × 20 × 10 mm$^3$ dimension by three stacking manners, respectively. The samples were irradiated from through-plane direction (TP) and in-plane direction (IP) with the same X-ray irradiance intensity, respectively.

Samples with dimension of 10 × 10 × 1 mm$^3$ were prepared for wide-angle X-ray diffraction (WAXS) to further explore the degree of orientation of graphene flakes. The 2D WAXS patterns of printed samples were obtained using a flat plate detector (Pilatus 300 K, 487 × 619 pixels with a pixel size of 172 μm). Each 2D WAXS pattern was acquired with an exposure time of 600 seconds. Fit2D and Matlab software were used to analyze the 2D WAXS patterns. All X-ray images were corrected for background scattering by subtracting contributions from air. The orientation parameter ($f$) of samples could be calculated by the Hermans' method, which was defined as:

$$f = \frac{3 <\cos^2\varphi - 1>}{2}$$

$$<\cos^2\varphi> = \frac{\int_0^{\pi/2} I(\varphi)\sin\varphi\cos^2\varphi \, d\varphi}{\int_0^{\pi/2} I(\varphi)\sin\varphi \, d\varphi}$$

The value of $f$ was between 0 (indicating random orientation) to 1 (indicating perfect orientation). I($\varphi$) represented the one-dimensional intensity distribution along the azimuthal angle ($\varphi$) of the (002) plane.

**UV-Vis-NIR Spectrophotometer Characterization.**

The G-TPU filament and C-TPU filament were dissolved in DMF solvent at a concentration of 2 mg/mL. The solutions were characterized using a UV-Vis-NIR spectrometer (UV-3600, Shimadzu, Japan) at the range of 300 – 1100 nm. Transmission spectra with range of 400 – 2000 nm were obtained from 20 × 20 × 1 mm³ films printed using G-TPU, C-TPU, and N-TPU filaments, respectively. These spectra were processed and analyzed using an ultraviolet-visible-near-infrared spectrophotometer (SOLID 3700, Shimadzu, Japan).

**Thermal Conductivity Measurements.**

For the measurement of G-TPU/N-TPU double-layer structure's anisotropic thermal conductivity ratio, the sample was prepared as **Figure S4a** and **S4b** to obtain the double-layer structure's TP and IP thermal conductivity, respectively. It should be noted that it was hardly possible to directly measure the IP thermal conductivity of G-TPU/N-TPU double-layer structure, therefore, we prepared the sample indicated in **Figure S4b** with the enlarged heat flux cross section for evaluation.

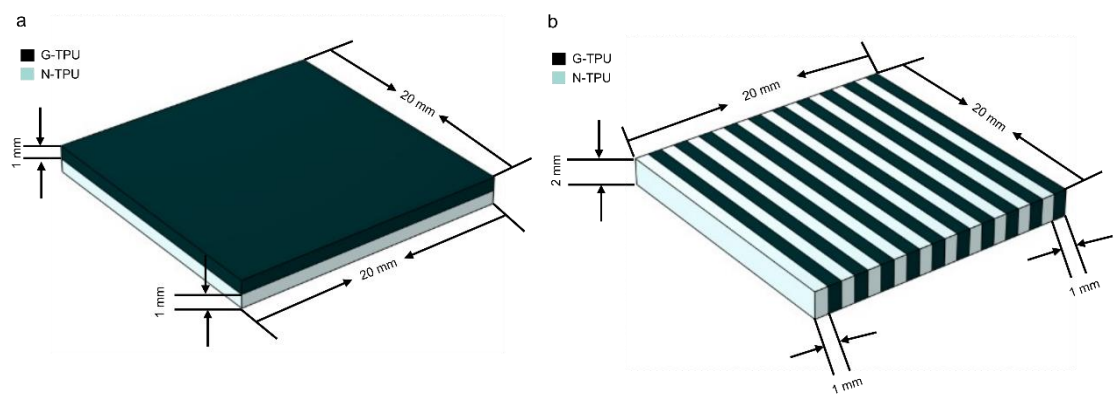

**Figure S4** Prepared sample for the thermal conductivity measurement of G-TPU/N-TPU double-layer structure in **(a)** TP and **(b)** IP direction with model dimensions indicated.

For the assessment of G-TPU layer's anisotropic thermal conductivity, the sample was prepared as a plate model with overall dimensions of 20 × 20 × 2 mm³ organized in a [0°/0°] or [0°/90°] stacking manners. As shown in **Figure S5**, by properly designing the layout of sample model, the anisotropic thermal conductivity of G-TPU layer in three different directions can be measured.

| Layout / Stacking Manners | | | |
|---|---|---|---|
| [0°/0°] | 4.54 W/(m·K) | 1.93 W/(m·K) | 0.77 W/(m·K) |
| [0°/90°] | 2.80 W/(m·K) | 2.80 W/(m·K) | 0.77 W/(m·K) |

**Figure S5** Anisotropic thermal conductivity of G-TPU sample in different layouts with [0°/0°] or [0°/90°] stacking manners, the red arrow indicated the thermal conductivity direction.

Thermal conductivity was calculated on the basis of the following equation: $TC = \alpha C_p \rho$,[2] where $\alpha$ is thermal diffusivity, $C_p$ is specific heat capacity, and $\rho$ is the density of the studied materials. Thermal diffusivity was acquired by laser flash method (LFA 457, NETZSCH, Germany). Specific heat capacity was obtained by a differential scanning calorimeter (DSC25, TA Instruments, America ) with a heating rate of 5 °C /min from − 10 to 50 °C. $\rho$ was measured using the water exclusion method based on Archimedes principle.

**Anisotropic Thermal Conductivity Test.**

The arrow shaped samples structures were fabricated using 30 wt% G-TPU with three distinct stacking manners ([0°/0°], [0°/90°] and [90°/90°]). Each sample featured a 30 wt% G-TPU layer with 0.5 mm thickness, and the structure was in overall dimensions of 30 × 20 × 0.5 mm³ (illustrated in **Figure S6**), specifically for thermal conductivity tests.

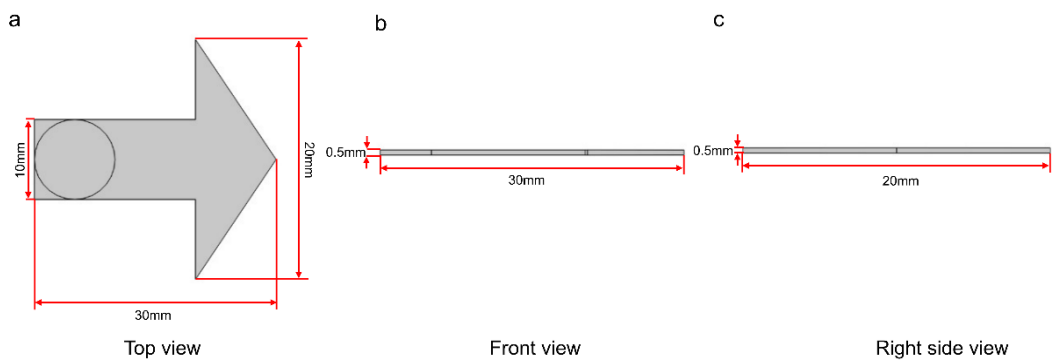

**Figure S6 (a)** Top view, **(b)** front view and **(c)** right view of the printed G-TPU layer arrow sample with model dimensions indicated.

These samples were horizontally placed on a platform and subjected to solar illumination. The light from the solar simulator was then focused through a magnifying glass achieving an illumination intensity of 0.8 W/cm², to specifically target the tail of the arrow samples. Shown in **Figure S7**, the infrared image and temperature distribution of the arrow sample's head were recorded every 60 s using an infrared camera (FLUKE TiS55+). The data were then imported into SmartView Classic 4.4 software for analysis.

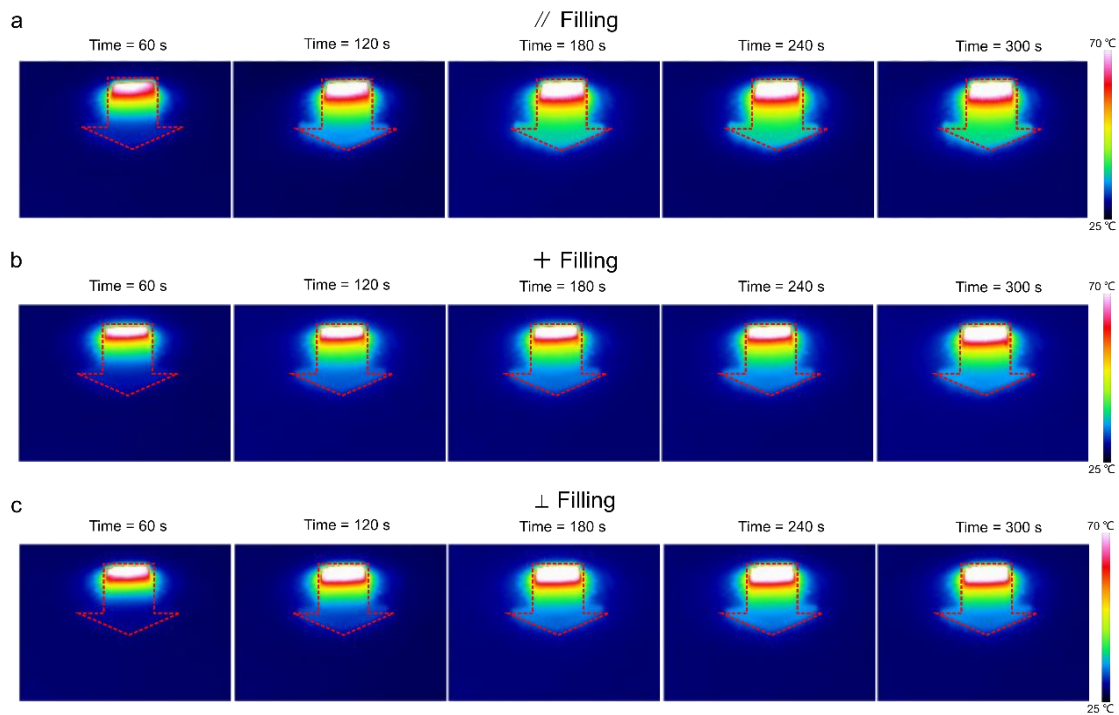

**Figure S7** Infrared image comparison of G-TPU layer arrow samples printed with different stacking manners: (a)

[0°/0°] stacking, (b) [0°/90°] stacking, (c) [90°/90°] stacking.

**Heat Retention Test.**

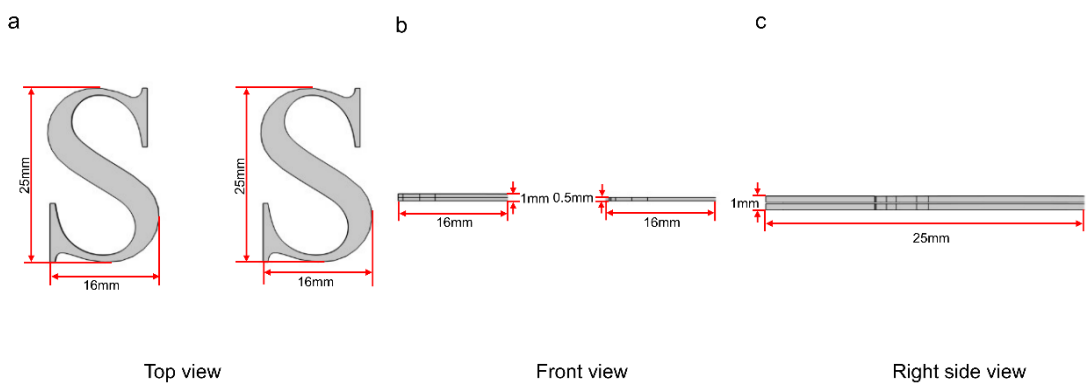

**Figure S8 (a)** Top view, **(b)** front view and **(c)** right view of the printed G-TPU/N-TPU double-layer letter (left) and G-TPU single-layer letter (right) sample with model dimensions indicated.

Two different structures were fabricated using a dual-nozzle 3D printer with [0°/90°] stacking manner for thermal insulation test. Specifically, the samples of G-TPU/N-TPU double-layer structures were printed with top layer of 0.5 mm 30 wt% G-TPU and bottom layer of 0.5 mm N-TPU (as shown in the left images of **Figure S8a** and **S8b**), and the G-TPU single-layer structures (as shown in the right images of **Figure S8a** and **S8b**) were printed in similar structure with thickness of 0.5 mm.

Both samples were heated at 60 °C in the oven until equilibrium and subsequently transferred into a room temperature environment. Infrared images and temperature distributions of the samples were recorded and documented using an infrared camera (FLUKE TiS55+).

**Photothermal Test.**

For photothermal IR imaging test, the letter samples of G-TPU/N-TPU double-layer structures were printed in similar structure with [0°/90°] stacking manner, consisting of 0.5 mm bottom layer of N-TPU and 0.5 mm top layer of 30 wt% G-TPU, measuring $16 \times 25 \times 1$ mm³ (shown in **Figure S9a** and **S9b left**) for conducting photothermal test.

For comparison, the letter samples of C-TPU single-layer structures were printed in similar structure with [0°/90°] stacking manner (shown in **Figure S9a** and **S9b right**) measuring 16 ×

25 × 1 mm³.

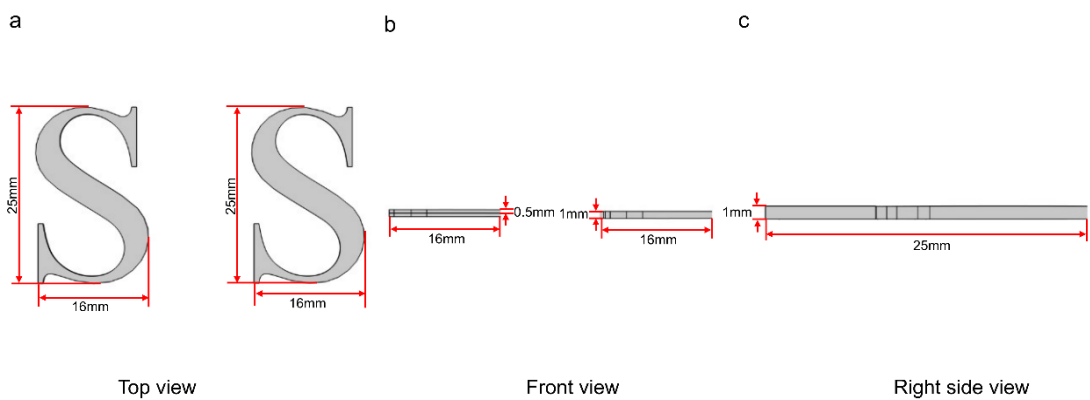

**Figure S9 (a)** Top view, **(b)** front view and **(c)** right view of the printed G-TPU/N-TPU double-layer letter (left) and C-TPU single-layer letter (right) sample with model dimensions indicated.

The photothermal test was conducted at an ambient temperature of approximately 22 °C. The printed letter samples were positioned on a board and irradiated using a solar simulator at an illumination intensity of 0.15 W/cm². An infrared camera (FLUKE TiS55+) recorded the samples every 10 seconds to capture infrared images and map the temperature distribution.

**Mechanical Tests.**

Tension tests were conducted on a universal testing machine (AGS-X2kN, Shimadzu, Japan) according to the ISO 37 standard. The thickness and length of the dumbbell-shaped samples were 2 mm and 75 mm, respectively. These samples were printed with different stacking manners, with a loading rate set at 50 mm/min. The modulus was calculated within the strain range of 1 – 5 %. Five samples of each composite type were tested, and the average tensile strength and elastic modulus were calculated.

Shear testing was performed according to ASTM C1292 using a universal testing machine (C45, MTS Systems Corporation, America). Three samples of each composite type were tested, and the average shear modulus was calculated.

The coefficient of friction (COF) was measured using a friction-abrasion testing machine (MFT-5000, Rtec Instruments, America) under air conditions. Steel balls (Φ = 6 mm) served as friction counter pairs, applied with a load of 2 N. The reciprocating amplitude was 10 mm at a frequency of 1 Hz.

Impact testing was conducted according to ASTM D3763 using a drop-weight impact testing machine (Amsler HIT 2000F, Zwick Roell, Germany). The drop-weight diameter was 16 mm, and the impact speed was approximately 3 mm/s.

**Numerical Simulations.**

The finite element method simulation was performed with the heat transfer module of COMSOL Multiphysics.

For anisotropic thermal conductivity simulation, G-TPU layer arrow was modeled according to the printing model file in **Figure S6**. Shown in **Figure 4g**, The parameters of the materials (listed in **Table S2**) were applied, and the boundary conditions were set as the convective heat flux (upper layer and side walls of the structure), thermal insulation (bottom layer of the structure) and illumination region with input power of 0.8 W/cm² in the pink region. The heat transfer module with coupled interfaces of solid and radiative beam in absorbing media was applied.

**Table S2** The main parameters of the materials used for numerical simulation

| Materials | 30 wt% G-TPU | | | N-TPU | C-TPU |
|---|---|---|---|---|---|
| Stacking Manners | [0°, 0°] | [0°, 90°] | [90°, 90°] | | |
| Thermal Conductivity in X Direction | 4.54 W/(m·K) | 2.80 W/(m·K) | 1.93 W/(m·K) | Isotropic, 0.24 W/(m·K) | Isotropic, 0.24 W/(m·K) |
| Thermal Conductivity in Y Direction | 1.93 W/(m·K) | 2.80 W/(m·K) | 4.54 W/(m·K) | | |
| Thermal Conductivity in Z Direction | 0.77 W/(m·K) | 0.77 W/(m·K) | 0.77 W/(m·K) | | |
| Heat Capacity | 1426 J/(kg·K) | | | 1500 J/(kg·K) | 1500 J/(kg·K) |
| Density | 1515 kg/m³ | | | 1200 kg/m³ | 1200 kg/m³ |

| Absorption Coefficient | 2.21 m$^{-1}$ | 0 m$^{-1}$ | 0.08 m$^{-1}$ |

Heat retention simulation was performed with the heat transfer module, the subjects were modeled as description in **Figure S8**, and the material's properties of each domain were defined as **Table S2**. And the boundary conditions were defined as thermal insulated bottom layers and natural convective cooling heat flux to the external environment with upper layers and walls. The overall structure was finely meshed, and then the computation was performed with 1,000,000 degrees of freedom.

The photothermal simulation was performed with the heat transfer module with coupled interfaces of solid and radiative beam in absorbing media. The subjects were modeled according to the sketch in **Figure S9**, and the material's properties of each domain were defined as **Table S2.** Notably, the absorption coefficient of 30 wt% G-TPU and N-TPU were estimated via quantifying their dissolved solution's optical density in the UV-Vis-NIR spectrum in **Figure 6h**. And the boundary conditions were defined as thermal insulated bottom layers, natural convective cooling heat flux to the external environment with upper layers and walls and the deposited beam power of 0.14 W/cm² in the upper layer. The overall structure was finely meshed, and then the computation was performed with 1,000,000 degrees of freedom.

**Molecular Dynamics Simulations.**

The molecular dynamics simulations were performed by Materials Studio. The TPU matrix of filament is of molecular weight of ~1,500,000. To simplify the computation, the box of N-TPU or G-TPU were modeled by packing the graphene flakes and repeating unit of TPU into a box to reach the density of ~1.3 g/cm³ that close to the experimental value. The repeating unit of TPU was described as one soft chain block that developed from octanediol ($C_8H_{18}O_2$) and diethelene glycol ($C_2H_6O_2$) and one hard chain that developed from methylenediphenyl diisocyanate ($C_{15}H_{10}N_2O_2$) and diethelene glycol ($C_2H_6O_2$). The geometric optimization was performed with Dmol3 module for energy minimization. The graphene flake was described as monolayer with dimension of 17 A × 19 A. The N-TPU box and G-TPU (30 wt% graphene) box with dimension of 28.9 A × 28.9 A × 9.7 A were built by packing desired composition to reach defined density with Amorphous Cell module. The built box was then assembled into Cu/N-TPU/Cu layer or Cu/G-TPU/Cu layer, and 100,000 cycles annealing steps were performed for stabilization. Then mechanical properties study and confined shear simulation were performed with assignment of COMPASS II forcefield.

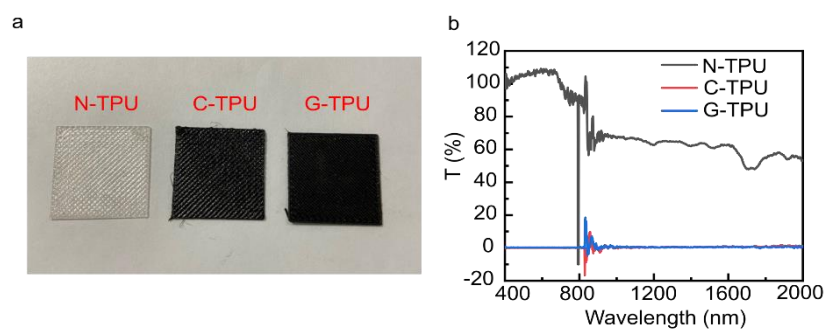

**Figure S10** Transmittance of N-TPU, C-TPU and 30 wt% G-TPU.

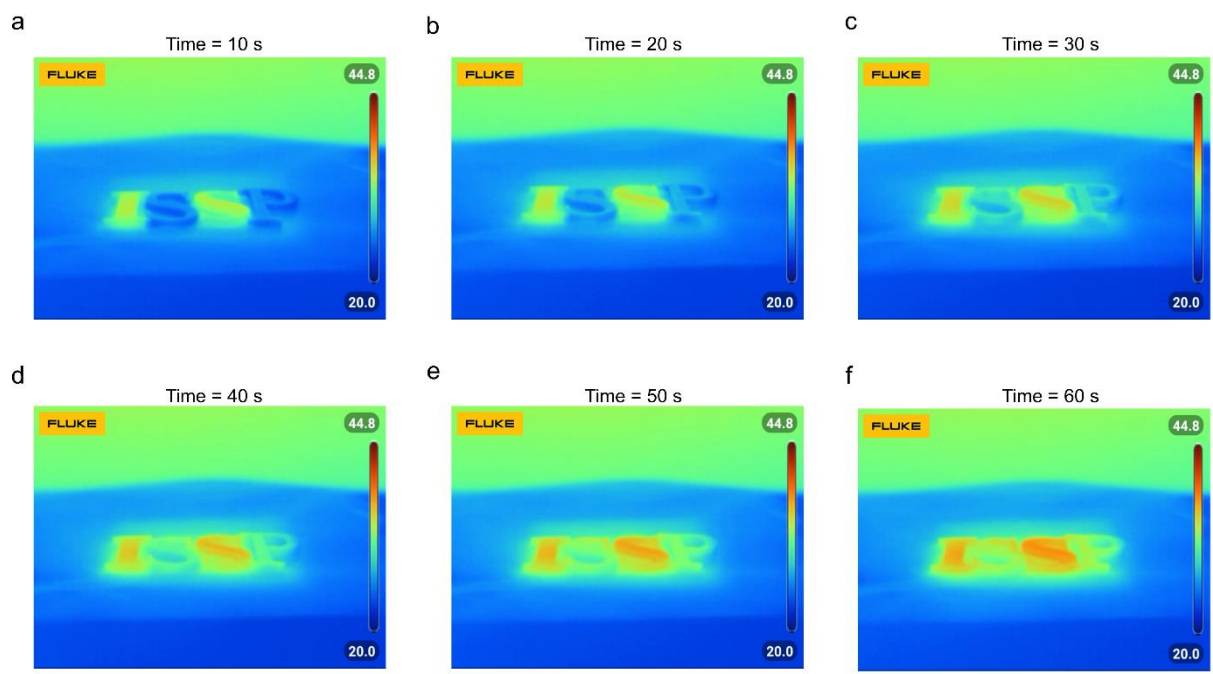

**Figure S11** Infrared photos of samples within 60 s under simulated sunlight irradiation with a light intensity of 0.15 W/cm². Letters "I" and "S" were printed as G-TPU/N-TPU double-layer structure, while letters "S" and "P" were printed with C-TPU filament of the same thickness to form the acronym "ISSP".